\documentclass[aip,jcp,reprint,noshowkeys,noshowpacs,groupedaddress]{revtex4-1}
\usepackage{textcomp} 
\usepackage{upgreek}
\usepackage{amsmath,amssymb,xfrac}
\usepackage{mathtools}
\usepackage{bm} 
\usepackage{graphicx}
\usepackage[caption=false]{subfig}
\usepackage{acronym} 
\usepackage{tabularx}
\makeatletter
\newcommand{\raisemath}[1]{\mathpalette{\raisem@th{#1}}}
\newcommand{\raisem@th}[3]{\raisebox{#1}{$#2#3$}}
\makeatother
\DeclareMathAlphabet{\mathsfit}{\encodingdefault}{\sfdefault}{m}{sl} 
\newcommand{\kB}{k_{\mathrm{B}}} 
\newcommand{\vc}[1]{{\mathbf{#1}}} 
\newcommand{\tz}[1]{\mathsf{#1}} 
\newcommand{\tzC}[1]{\mathsfit{#1}} 
\newcommand{\tsc}[1]{\text{\textsc{#1}}} 
\newcommand{\I}{\mathcal{I}} 
\newcommand{\Bo}{\mathcal{B}} 
\newcommand{\dB}{{\partial\Bo}} 
\newcommand{\dV}{{\mathrm{d}V}} 
\newcommand{\dS}{{\mathrm{d}S}} 
\newcommand{\dO}{{\mathrm{d}\Omega}} 
\newcommand{\dx}{{\mathrm{d}x}} 
\newcommand{\s}{\bm{\upsigma}} 
\newcommand{\e}{\bm{\upvarepsilon}} 
\newcommand{\sC}{\sigma} 
\newcommand{\eC}{\varepsilon} 
\newcommand{\eT}{\e^\tsc{t}} 
\newcommand{\sT}{\s^\tsc{t}} 
\newcommand{\eTC}{\eC^\tsc{t}} 
\newcommand{\sTC}{\sC^\tsc{t}} 
\newcommand{\eP}{\e^\tsc{p}} 
\newcommand{\sP}{\s^\tsc{p}} %
\newcommand{\sPC}{\sC^\tsc{p}} %
\newcommand{\eS}{\e^\tsc{s}} 
\newcommand{\sS}{\s^\tsc{s}} 
\newcommand{\eSC}{\eC^\tsc{s}} 
\newcommand{\sSC}{\sC^\tsc{s}} 
\newcommand{\Lt}[2]{\tz{L}\!\left[{#1}\,,\,{#2}\right]} 
\newcommand{\Dt}[2]{\tz{D}\!\left[{#1}\,,\,{#2}\right]}
\newcommand{\Bm}{\kappa} 
\newcommand{\mI}{\mu_0} 
\newcommand{\lI}{\lambda_0} 
 
\newcommand{\nI}{\nu_0}
\newcommand{\kI}{\Bm_0}
\newcommand{\cI}{\tz{C}_0}

\newcommand{\sus}{\tz{X}} 
\newcommand{\susC}{\tzC{X}}
\newcommand{\susS}{\tz{X}^\tsc{s}} 
\newcommand{\susSC}{\tzC{X}^\tsc{s}}
\newcommand{\smr}{(\mu/\mu_0)} 
\newcommand{\MR}{\tz{Y}} 

\newcommand{\mM}{\mu}
\newcommand{\nM}{\nu}
\newcommand{\kM}{\Bm}
\newcommand{\cM}{\tz{C}}

\newcommand{\fsn}{\bm{\upeta}} 
\newcommand{\fsnC}{\eta} 
\newcommand{\fsC}{\omega} 
\newcommand{\fs}{\bm{\upomega}} 
\newcommand{\ut}{\bm{\uppsi}} 
\newcommand{\utC}{\psi} 
\newcommand{\pE}{\mathcal{E}} 
\newcommand{\fE}{\mathcal{F}} 
\newcommand{\Ham}{\mathcal{H}} 
\newcommand{\half}{\frac{1}{2}}
\newcommand{\lp}{\left(}
\newcommand{\rp}{\right)}
\newcommand{\lb}{\left[}
\newcommand{\rb}{\right]}
\newcommand{\la}{\langle}
\newcommand{\ra}{\rangle}
\newcommand{\rf}[1]{(\ref{#1})}

\newcommand{\ofR}{({\mathbf{r}})} 
\newcommand{\inc}{{\operatorname{inc}\,}}
\newcommand{\rot}{{\operatorname{rot}\,}}

\newcommand{\QQText}[1]{\text{\qquad{#1}\qquad}}
\providecommand{\abs}[1]{\lvert#1\rvert}

\newcommand{\rmd}{\mathrm{d}}
\newcommand{\gA}{{\bm{\upgamma}^\tsc{a}}}
\newcommand{\gAC}{{\gamma}^\tsc{a}}

\newcommand{\sigC}[1]{\sigma^{(i)}}

\newcommand{\spinC}[1]{\utC^{(i)}}
\newacro{RFOT}[RFOT]{random first order transition} 
\newacro{CRR}[CRR]{cooperatively rearranging region} 
\newacro{MCT}[MCT]{mode-coupling theory}
\newacro{DFT}[DFT]{density functional theory}
\newacro{TLS}[TLS]{two-level systems}
\newacro{MF}[MF]{mean field}
\newacro{BC}[BC]{boundary conditions}
\newacro{MC}[MC]{Monte Carlo}
\newacro{RCP}[RCP]{random close packed}
\newacro{PES}[PES]{potential energy surface}
\newacro{rhs}[rhs]{right-hand side}
\newacro{lhs}[lhs]{left-hand side}
\newacro{MD}[MD]{Molecular Dynamics}
\newcommand{\thetaF}{\theta^\tsc{f}} 
\newcommand{\thetaS}{\theta^\tsc{s}} 
\newcommand{\thetaG}{\theta^\tsc{g}} 
\newcommand{\pEF}{\pE^\tsc{f}}
\newcommand{\pES}{\pE^\tsc{s}}
\newcommand{\pEG}{\pE^\tsc{g}}
\newcommand{\prtl}{\partial} 
\newcommand{\svibr}{{\mbox{\scriptsize vibr}}}
\begin{document}

\newlength{\figurewidth}
\setlength{\figurewidth}{.6\columnwidth}

\title{Self-consistent elastic continuum theory of degenerate,
  equilibrium aperiodic solids} \author{Dmytro Bevzenko}
\affiliation{Department of Chemistry, University of Houston, Houston,
  TX 77204-5003}

\author{Vassiliy Lubchenko}
\email[]{vas@uh.edu}
\affiliation{Department of Chemistry, 
             University of Houston, Houston, TX 77204-5003}
\affiliation{Department of Physics, 
             University of Houston, Houston, TX 77204-5005}
\date{\today}
\begin{abstract}
  We show that the vibrational response of a glassy liquid at finite
  frequencies can be described by continuum mechanics despite the vast
  degeneracy of the vibrational ground state; standard continuum
  elasticity assumes a unique ground state. The effective elastic
  constants are determined by the bare elastic constants of individual
  free energy minima of the liquid, the magnitude of built-in stress,
  and temperature, analogously to how the dielectric response of a
  polar liquid is determined by the dipole moment of the constituent
  molecules and temperature. In contrast with the dielectric
  constant---which is enhanced by adding polar molecules to the
  system---the elastic constants are down-renormalized by the
  relaxation of the built-in stress.  The renormalization flow of the
  elastic constants has three fixed points, two of which are trivial
  and correspond to the uniform liquid state and an infinitely
  compressible solid respectively.  There is also a nontrivial fixed
  point at the Poisson ratio equal to 1/5, which corresponds to an
  isospin-like degeneracy between shear and uniform deformation.  The
  present description predicts a discontinuous jump in the (finite
  frequency) shear modulus at the crossover from collisional to
  activated transport, consistent with the RFOT theory.
\end{abstract}
\maketitle
\section{Introduction}\label{sec:intro}

In the absence of kinetic access to a crystalline or partially ordered
state, a liquid can be equilibrated even below the fusion
temperature. If such a liquid is sufficiently pressurized and/or
cooled, it undergoes a crossover from largely collisional to activated
transport,\cite{LW_soft, LW_Wiley} whereby long-lived aperiodic
structures begin to form; these can be seen directly by neutron
scattering.\cite{MezeiRussina} The crossover is manifested
thermodynamically as a breaking of the translational symmetry upon
which the particle density profile is no longer uniform but consists
of disparate, narrow peaks.\cite{dens_F1} For example, in ordinary,
chemically-bonded liquids the crossover takes place at viscosity
values around 10~Ps or, equivalently, when the $\alpha$-relaxation
time is about three orders of magnitude longer than the vibrational
relaxation time: $\tau_\alpha \simeq 10^3 \tau_\svibr$.\cite{LW_soft,
  RL_Tcr} The crossover to activated transport can occur either below
or above the fusion temperature, depending on the liquid's
fragility.\cite{LW_soft} In the latter case, the liquid is technically
supercooled. For generality, we will use the term ``glassy'' for a
liquid below the crossover---but above the glass transition---since
the glass transition is always preceded by the crossover in ordinary
liquids.

As worked out in the random first order transition (RFOT) theory,
particles move below the crossover via local activated
reconfigurations between distinct aperiodic free energy
minima,\cite{KTW, XW} see Ref.~\onlinecite{LW_ARPC} for a review.
These reconfigurations are responsible for the $\alpha$-relaxation.
They involve several hundred atoms near the glass transition; the
corresponding cooperativity length scale $\xi$ is numerically $2-4$~nm
in actual substances,\cite{XW, LW} consistent with
observation.\cite{GruebeleSurface, Spiess, RusselIsraeloff,
  CiceroneEdiger} The cooperative reconfigurations are driven by the
multiplicity of the distinct aperiodic free-energy minima, whose
log-number is called the configurational entropy. The configurational
entropy is inherently connected and numerically close to the excess
liquid entropy relative to the corresponding crystal; this excess
entropy can be inferred from experiment.\cite{Capaccioli, RWLbarrier}

The activated reconfigurations restore the ergodicity and dictate that
the zero-frequency modulus be zero. Despite this liquid-like response
at the very lowest frequencies, the material exhibits {\em elastic}
response at non-zero frequencies.  The vibrational response of
supercooled liquids, at these frequencies, apparently obeys standard
continuum mechanics and can be measured, for instance, by Brillouin
scattering.\cite{RL_Tcr} Yet continuum mechanics assumes at the onset
that there is a unique vibrational ground state. Under this
assumption, the particle identities in the ground state and in a
vibrationally excitated state can be strictly matched thus allowing
one to define local displacement $\vc{u}$ unambiguously. In contrast,
a liquid in the activated transport regime---as it would be near its
glass transition---is a mosaic of aperiodic structures each
corresponding locally to distinct, individual minima of the free
energy;\cite{XW} the built-in stress at the physical boundaries
between the structures, due to the mutual mismatch, cannot be removed
by elastic deformation. While vibrational excitations within
individual minima are well defined, this is not so for the actual
liquid, because the structure relaxes on a finite time scale. In fact,
a liquid of volume $V$ will experience local relaxation roughly once
per time $\tau_\alpha \xi^3/V$.\cite{Lthinning} Thus the larger the
region, in which one considers vibrational excitations such as sound
waves, the more ambiguous it is to define a vibrational ground state.

Here we determine the vibrational response of such an equilibrium,
degenerate aperiodic solid starting from the elastic properties of
{\em individual} aperiodic free energy minima.  We show that the
question is in many ways analogous to the problem of determination of
the dielectric response of a fluid given the dipole moment and
polarizability of the constituent molecules.\cite{Onsager1936,
  debye12, debye29, kirkwood39} The role of the permanent dipoles is
played here by the built-in mechanical stress, a tensorial quantity.
Even when mechanically stable, all solids are inherently
stressed:\cite{Alexander1998} For instance, in a bulk periodic
crystal, the bond lengths differ from their values in very small
clusters made of the same material.  Crystal surfaces are often
reconstructed.\cite{Bienfait, SurfRough} These are rather trivial
examples in that the stress can be removed by deformation without
breaking bonds; this simple kind of stress not classified as built-in.
Much more interesting are strains arising in the presence of
vacancies/interstitials, dislocations, or disclinations. These sources
of strain {\em cannot} be removed without breaking bonds.  A simple
but key signature of built-in stress that it cannot have an
arbitrarily small magnitude; the magnitude must be {\em finite}.
Conversely, stress of arbitrarily small magnitude can be removed by
elastic deformation.

In glassy solids, local stresses are mutually frustrating and lead to
structural {\em degeneracy}, which is manifested thermodynamically as
the configurational entropy, as mentioned.  The concentration of the
stressed regions is inherently $1/\xi^3$, where $\xi$ is the
volumetric size of the cooperatively rearranging region during
$\alpha$-relaxation.  A similar example of such frustration
constructed theoretically is that arising in icosahedral
order;\cite{PhysRevB.32.1480} the corresponding free energy landscape
is consistent with the predictions of the RFOT
theory.\cite{Nussinov_nonAb} Inherent stresses arising in solids owing
to aperiodicity have been discussed previously in
Refs.~\onlinecite{0305-4608-12-10-010, Srolovitz1981,
  0295-5075-104-5-56001}.

The picture of a supercooled liquid as a stressed degenerate continuum
emerges in the analysis by the present authors\cite{BL_6Spin} (BL),
which is complementary to the RFOT theory in that it considers a
non-degenerate, stable solid---not the uniform liquid---as the
reference state for building the glassy state. (Presumably, such a
non-degenerate solid is ordinarily periodic in 3D.) In the BL
construct, one splits the total deformation tensor:\cite{LLelast}
\begin{equation}
\label{eq:def}
\eC_{ij} \equiv \half\lp\frac{\partial u_i}{\partial x_j}+\frac{\partial
  u_j}{\partial x_i}\rp  \equiv \half\lp u_{i,j}+u_{j,i}\rp 
\end{equation}
into a sum of a small-$k$ (long-wavelength) contribution $\eC_{ij}^<$
and large-$k$ (short-wavelength) contribution $\eC_{ij}^>$: $\eC_{ij}
= \eC_{ij}^< + \eC_{ij}^>$. Upon denoting the short-wavelength part as
$\fsnC_{ij} \equiv \eC_{ij}^>$, the usual vibrational free
energy\cite{LLelast} reads:
\begin{equation} \label{eq:Anzatz}
  \fE=\half\int\dV\lp\e^<+\fsn\rp\cI\lp\e^<+\fsn\rp,
\end{equation}
where $\cI$ is the elastic moduli tensor.  Subsequently, one fixes the
magnitude of the short-wavelength stress:
\begin{equation} \label{eq:op:local}
  \fsn\ofR\cI\fsn\ofR=g^2\ofR.
\end{equation}
thus (artificially) making it built-in.  With this constraint, the
simple model from Eq.~\eqref{eq:Anzatz} becomes strongly non-linear.
We associate the lengthscale below which the stress cannot relax with
the size $a$ of the chemically-rigid molecular unit, or
``bead.''\cite{LW_soft, BL_6Spin} Conversely, the elastic degrees of
freedom $\eC_{ij}^<$ are essentially phonons with $k < \pi/a$.

We have shown that given a large enough magnitude $g$ of built-in
stress, there emerges self-consistently a metastable,
structurally-degenerate aperiodic state separated by a nucleation
barrier from the stable, unique reference state. In the simplest
treatment, one finds that the structural degeneracy of a supercooled
liquid maps onto the set of mutual orientations of an assembly of
six-component Heisenberg spins on a fixed lattice with anisotropic
interactions. The six components reflect the number of independent
entries of the deformation tensor from Eq.~\eqref{eq:def}.  One can
make parallels between the BL picture and that by Yan et
al.\cite{Yan16042013}, in whose model the degeneracy is built-in by
assuming individual bonds can switch between two alternative lengths,
where the switching is controlled by an Ising-like variable.

In terms of the aforementioned analogy with the dielectric response,
the {\em stable} vibrational ground state---i.e., elastic medium
without built-in stress---corresponds to vacuum, while the sources of
stress correspond to molecular dipoles. As in the dielectric case, the
interaction between the sources of stress scales with the distance $r$
as $1/r^3$, although it is now of more complicated, tensorial form.

There are several, distinct motivations for the present
calculation. The most immediate motivation is to connect the
characteristics of local stress to the elastic properties of the
solid, much like Onsager determined the dielectric response of a
liquid using the dipole moment and polarizability of individual
molecules as the microscopic input. To quantify the renormalization of
the elastic moduli---and especially their decrease upon approaching
the cross-over from below---is essential for building a theory of the
glass transition.\cite{PhysRevLett.105.015504, Yan16042013,
  7644758720120607}

A distinct motivation is to accomplish the BL programme of detailed
characterization of the activated dynamics in liquids via the
6-component spin model, which has certain advantages over direct
simulation of liquids: The spins are not subject to collisional
effects that represent a significant source of slowing down in liquid
simulations. The spin model has a significantly smaller number of
degrees of freedom than the corresponding liquid since the purely
vibrational modes $\e$ can be integrated out.  In addition, the spins
are situated on a {\em fixed} lattice, making it easier to define an
order parameter for activated reconfigurations, so that configurations
can be distinguished based on the orientations of the 6-spins. An
explicit advantage of the elasticity-based approach of BL is that the
complicated inter-atomic forces enter the description only through
very few parameters. In the most minimal description, this set of
parameters includes only the compressibility, shear modulus, and bead
size. Conversely, the explicit functional form of the many-body forces
in actual materials is simply unavailable even though simplified,
effective potentials, such as the BKS model~\cite{PhysRevLett.64.1955}
of amorphous silica, have been reasonably successful in reproducing
several material properties.  Incidentally, direct simulations of
actual liquids still remain excessively computationally costly. Only
for simple systems, such as Lennard-Jones or hard sphere mixtures, the
onset of activated transport seems to have been reached in simulation,
see Ref.~\onlinecite{PhysRevLett.112.097801} and references therein.

One of the most challenging aspects of the BL program is that the
spin-spin interaction scales as $1/r^3$ and thus is much longer-range
than ordinary molecular interactions; this potentially leads to
artifacts in simulations due to finite-size effects. For instance,
imposing periodic boundary conditions on models with such long-range
interactions will likely produce excessive finite-size effects.
Indeed, simulations of dipolar systems on periodic lattices have
produced ordered states.\cite{ayton95} To avoid such artifacts, one
may employ a different type of boundary conditions, in which the spins
inside a compact region are treated explicitly, while the outside
spins are approximated as an elastic continuum. This is in direct
analogy with the Onsager cavity construction,\cite{Onsager1936} except
here one treats the number of particles inside the cavity as a
flexible parameter; the Onsager limit is achieved in the limit of one
spin per cavity. The cavity construction is often used in computer
simulations of polar liquids.\cite{Tomasi1994} Additionally, imposing
the self-consistency in the determination of the elastic response
lends further support to the BL picture, as the latter is not fully
self-contained: The stabilization of the aperiodic phase stemming from
steric repulsion, mentioned earlier, is not explicitly treated in the
present version of the BL formalism, but is assumed. Finally,
achieving the self-consistency using a continuum treatment alleviates
concerns about the ultraviolet behavior of the BL model, in which
local sources of built-in stress are approximated as point-like
objects, while their mutual spacing enters through the ultraviolet
cut-off in phonon sums.

Last, but not least, this work addresses the fundamental challenge of
developing continuum mechanics for a medium that has a vastly
degenerate ground state. Ordinary theory of elasticity~\cite{LLelast}
simply assumes a unique reference state exists. All excitations in the
latter theory are diffeomorphisms, i.e., combinations of stretches and
contractions. The resulting states are all equivalent from the
viewpoint of differential geometry since they have the same
connectivity. In chemical language, no bonds can be broken or made
during such elastic deformation. The above notions can be formalized
as follows. The energy of an elastic deformation can only depend on
the spatial {\em derivative} of the actual atomic displacement
$\vc{u}$ since this energy does not depend on the absolute location of
the body in space. Thus in the lowest order, the deformation is
described by a (symmetric) tensor from Eq.~\eqref{eq:def}
which has six independent components and thus potentially over-defines
the actual particle displacement, which has only three independent
components. The conventional continuum mechanics adopts a specific
condition on the $\eC_{ij}$ tensor that turns out to supply exactly
three constraints.  This condition insures that the integration of the
deformation tensor $\eC_{ij}$---with the aim of computing the actual
displacement $\vc{u}$---gives the same result regardless of the
contour of integration. In chemical language, this is equivalent to
requiring that no bonds are broken during deformations. By the
Saint-Venant theorem, see e.g. Ref.~\onlinecite{lovett91}, this can be
achieved, if the so called ``incompatibility'' tensor is identically
zero:
\begin{equation} \label{eq:SaintVenant} \lp\inc\e\rp_{ij} \equiv
  -\epsilon_{ikl} \epsilon_{jmn} \eC_{ln, \, km} =0,
\end{equation}
where $\epsilon_{ijk}$ is the Levi-Civita symbol. Throughout, we imply
summation with respect to doubly-repeated indices. Given a deformation
tensor $\eC_{ij}$ that satisfies constraint (\ref{eq:SaintVenant}),
the atomic displacement $\vc{u}$ can be unambiguously computed using
the Kirchhoff-Ces\`aro-Volterra formula.\cite{Teodosiu1982, lovett91}
Condition (\ref{eq:SaintVenant}) is analogous to the constraint one
imposes in electrodynamics (in the absence of charges) that the
electric field be rotor-free: $\bm{\nabla}\times\vc{E} = 0$. Only under
such circumstances can the electric field be expressed as the gradient
of a single-valued, scalar field; this is needed to make the energy of
an electric charge subject to electric field a well defined,
single-valued function of the coordinate. Note that the existence of a
unique reference state for the continuum mechanics is analogous to
stipulating that vacuum be unique in electrodynamics.

The differential-geometric formulation of continuum
mechanics~\cite{kroner92} generalizes the defect-free description
corresponding to Eq.~(\ref{eq:SaintVenant}) to more complicated
situations when dislocations and vacancies/interstitials are present,
by introducing {\em torsion} and {\em nonmetricity} respectively. Thus
one tacitly assumes there is an underlying Bravais lattice in the
continuous description. Applicability of such description to glassy
systems is far from certain however. On the one hand, there is no
underlying Bravais lattice in a supercooled liquid or glass. At the
same time, the coordination varies spatially.  Consequently,
describing the space itself, let alone potential defects in the space,
by continuum methods becomes ambiguous.  Generally, defining defects
in a disordered medium is ambiguous, too: As emphasized in
Refs.~\onlinecite{LW_RMP, ZL_JCP, ZLMicro1, ZLMicro2}, supercooled
liquids or glasses cannot be regarded as defected versions of crystal
since the crystal portion of the phase space is not accessible to the
system. Consistent with these notions, Cammarota and
Biroli~\cite{0295-5075-98-3-36005} argued there is no static pattern
corresponding to the metastability of a supercooled liquid with
respect to local reconfiguration between alternative free energy
minima.  Thus the lengthscale corresponding to those stress patterns
generally must be---and has been~\cite{GruebeleSurface, Spiess,
  RusselIsraeloff, CiceroneEdiger}---determined dynamically.  It is
not clear at present whether the local free energy excess due to
built-in {\em stress} in glassy liquids can be measured by linear
spectroscopy. Still, note that in one family of glasses,
viz. chalcogenide alloys, the stressed regions have an electronic
signature in the form of midgap electronic states~\cite{ZL_JCP,
  ZLMicro2} that can be detected by essentially linear
means.~\cite{BiegelsenStreet, PhysRevLett.42.118} In addition, the
amount of built-in stress may be modified by varying the speed of
quenching or as a result of polymerization below the glass transition,
leading to a change in vibrational properties of the
glass.~\cite{doi:10.1021/jp4054742}

The notions of the structural degeneracy and the resulting steady
structural reconfiguration between alternative aperiodic structures
are key to the present work. A fully {\em stable} lattice---periodic
or aperiodic---has a unique vibrational ground state, in contrast with
actual glassy liquids that are prevented from crystallization.  Even
though plane waves are no longer vibrational eigen-modes in a stable
aperiodic lattice, there is no ambiguity in defining an elastic
response down to zero frequencies. Far from simple, the vibrational
response of stable aperiodic lattices generally includes non-affine
displacements,~\cite{PhysRevLett.97.055501} which {\em also} violate
the Saint-Venant compatibility condition
(\ref{eq:SaintVenant}).~\cite{0953-8984-26-1-015007} Local elastic
response in aperiodic lattices is generally spatially
inhomogeneous;~\cite{6078032020110601, 0295-5075-104-5-56001,
  PhysRevB.79.060201} the distribution has been argued to cause
down-renormalization of the bulk elastic
constants.~\cite{0295-5075-73-6-892, Marruzzo2013}

The present theory of elasticity of equilibrium aperiodic solids, such
as supercooled liquids, is based on the notions of structural
degeneracy and built-in stress, not structural inhomogeneity per
se. The article works out the resulting microscopic picture in the
following logical sequence: In Section~\ref{sec:dielectric}, we
briefly review the theory of dielectrics, which relates the
expectation value of local polarization to the bulk dielectric
response of the material. There we also review Onsager's construction
for determining the local polarization and the effective dielectric
constant of the liquid self-consistently, based on the dipole moment
of individual molecules. Section~\ref{sec:symm-betw-electr}
demonstrates that the type of uniformly distributed built-in stress
characteristic of glassy liquids is analogous in several ways to
molecular dipoles in an equilibrated fluid. Alongside, the analogy
between continuum electrodynamics and mechanics is explained and
elements of tensor algebra that greatly facilitate the analysis of the
elastic case are reviewed. In Section~\ref{sec:elasticity:general}, we
make a connection between the expectation value of the built-in stress
and renormalization of the elastic constants. In
Section~\ref{sec:cavity:main}, we compute the interaction between
local sources of built-in stress, which is the analog of the
dipole-dipole interaction in
electrodynamics. Section~\ref{sec:cavity:gen} works out the
generalized cavity construction for elasticity. We obtain formal
expressions for the vibrational response of a degenerate, equilibrium
aperiodic solid, in which a compact subset of local sources of
built-in stress are treated explicitly while its environment is
approximated as an elastic continuum with effective elastic constants.
Section~\ref{sec:cavity:softening} determines the bulk elastic
response of such a solid approximately for three specific
implementations of the built-in stress. In all cases, the elastic
constants are {\em down}-renormalized owing to the built-in stress in
contrast with the dielectric case, in which the dielectric constant
can only be enhanced by molecular dipoles. In addition to the trivial
fixed points to the elastic renormalization---which correspond to the
uniform liquid and infinitely compressible solid---a special value of
the Poisson ratio, $\nu = 1/5$, emerges as a non-trivial fixed point
that corresponds to a special degeneracy between pure uniform and
shear deformations.

The first implementation of the built-in stress is closest in spirit
to the Onsager approximation and amounts to a source of built-in
stress directly in contact with the effective elastic medium. We
establish that there is a limiting value to the built-in stress past
which the mechanical stability limit of the aperiodic solid is
reached. We also find self-consistently that a uniform liquid cannot
sustain built-in stress.  The second implementation is appropriate for
realization of the BL program in which an arbitrarily large, compact
subset of the sources are treated explicitly while the environment is
approximated as an elastic continuum. The third implementation is a
systematically worked-out analog of how the built-in stress was set up
in the original BL paper.\cite{BL_6Spin} Here we find that the $\nu =
1/5$ fixed point is {\em repulsive}, in contrast with the first two
cases. This repulsive fixed point is consistent with the critical
point at $\nu = 1/5$ found in the mean-field limit of the BL
model. The corresponding continuous transition separates two
relatively distinct regimes in which a supercooled liquid can be
viewed as a frozen-in stress pattern corresponding to largely uniform
dilation/compression and shear respectively. In all three
implementations, we observe that the transition between the uniform
liquid and the degenerate, aperiodic crystal is discontinuous,
consistent with the RFOT theory. In the final
Section~\ref{sec:discussion}, we discuss and summarize the present
findings.

\section{Review of the cavity construction for polar liquids}
\label{sec:dielectric}

The present argument for determination of the mechanical response of
an aperiodic solid, as a degenerate collection of sources of stress,
is relatively complex mathematically, partially because of the
tensorial character of mechanical deformation. It seems most
profitable to present this argument by analogy with the simpler
calculation of the dielectric response of polar liquids, which are
characterized by a multiplicity of distinct configurations of the
molecular dipoles.

Consider a dielectric liquid with susceptibility $\epsilon$ and assume
that chemically inert, polar molecules are dissolved in the liquid at
a low concentration $c$. We label the magnitude of the permanent
dipole moment of the solute molecules by $d$ and neglect their
polarizability, since we will not be considering the elastic analog of
the polarizability in what follows. (``Elastic polarizability'' is
usually neglected in treatments of elastic defects.\cite{kroner60,
  Dederichs1978, Puls2012}) Our task is to determine the effective
dielectric constant $\epsilon'$ of the solution self-consistently.
Note we set up the dielectric problem a bit differently from the
conventional procedure, which fixes the bare dielectric susceptibility
in the absence of solute at its value in vacuum, whereby $\epsilon =
1$.

By definition, the local value of the electric displacement in the
solution is\cite{LLcont}
\begin{equation}
  \label{eq:def:LocalDispl}
  \vc{D}=\vc{E}+4\pi\lp\vc{P}^\tsc{b}+\vc{P}^\tsc{d}\rp, 
\end{equation}
where $\vc{E}$ is the local value of the electric field and the total
polarization is the sum of two components: the polarization
$\vc{P}^\tsc{b}$ of the bare solvent and the polarization
$\vc{P}^\tsc{d}$ of the solute. Since the dependence of
$\vc{P}^\tsc{b}$ on the electric field is known,
$\lp\epsilon-1\rp\vc{E}=4\pi\vc{P}^\tsc{b}$, it can be excluded from
Eq.~\eqref{eq:def:LocalDispl} to yield
\begin{equation}
  \label{eq:LocalDispl:noPb}
  \vc{D}=\epsilon\vc{E}+4\pi\vc{P}^\tsc{d}.
\end{equation}

The total dielectric constant $\epsilon^\prime$ of the solution can be
defined as the proportionality coefficient between the volume averages
of $\vc{E}$ and $\vc{D}$, similarly to how the effective dielectric
constant of a mixture is defined,\cite{LLcont}
\begin{equation} \label{eq:av:ee2}
  \overline{\vc{D}}=\epsilon^\prime\,\overline{\vc{E}}
  \QQText{or}\epsilon^\prime\,\overline{\vc{E}}=
  \epsilon\overline{\vc{E}}+4\pi\overline{\vc{P}}\,\!^\tsc{d},
\end{equation}
where the averaging is done over a volume containing an appreciable
number of solute molecules. As suggested by Eq.~\eqref{eq:av:ee2},
$\overline{\vc{P}}\,\!^\tsc{d}$ is a function of the mean field
$\overline{\vc{E}}$ only. In the linear-response regime, we obtain
\begin{equation} \label{eq:dipole:P:expansion2}
  \overline{P}\,\!^\tsc{d}_i =
  \chi_{ij}\overline{E}_j,
\end{equation}
where
\begin{equation} \label{eq:dipole:Chi} \chi_{ij} \equiv \frac{\partial
    \overline{P}\,\!^\tsc{d}_i}{\partial{\overline{E}}_j}
  \Bigg\vert_{\overline{\vc{E}}=0}
\end{equation}
is the static isothermal response function of the solute as dissolved
in the solvent. We have used that in equilibrium,
$\overline{\vc{P}}\,\!^\tsc{d}=0$ in the absence of external
field. Substituting Eq.~\eqref{eq:dipole:P:expansion2} into
Eq.~\eqref{eq:av:ee2} one obtains the following relation between the
bare and effective dielectric constants:
\begin{equation}
  \label{eq:dipole:ee3}
  \epsilon^\prime = \epsilon \delta_{ij}+4\pi\chi_{ij}.
\end{equation}
To calculate the susceptibility $\chi_{ij}$ we must use a specific
model for dipole dynamics in the solution. At high temperatures, a
good approximation is afforded by the Onsager cavity
construction.\cite{Onsager1936} Assuming the solute concentration is
$c$, the polarization density is, approximately,
\begin{equation}\label{eq:onsager:cavity:thepoint}
\overline{\vc{P}}\,\!^\tsc{d}\approx c\langle\mathbf{d}\rangle,
\end{equation}
where $\langle\mathbf{d}\rangle$ is the thermally averaged value of an
individual molecular dipole.  We treat an individual, chosen molecular
dipole explicitly while approximating the response of the rest of the
dipoles to the motions of the chosen dipole by the response of a
dielectric continuum with an effective dielectric constant
$\epsilon'$. The chosen dipole is placed, by construction, in the
center of a spherical cavity of radius $r_0 = (3/4 \pi c)^{1/3}$.  The
medium inside the cavity is still characterized by the bare dielectric
constant $\epsilon$.

The inhomogeneity in the local dielectric response due to the cavity
does not, on average, modify the electric displacement $\vc{D}$,
since the latter is determined by the charge distribution {\em
  outside} the sample. We assume, in a mean-field fashion, that the
displacement is in fact spatially homogeneous: $\vc{D} =
\overline{\vc{D}}$. Consequently, the electric field far away from the
cavity is also homogeneous and, by Eq.~\eqref{eq:av:ee2}, is equal to
the mean field $\overline{\vc{E}}$,
\begin{equation}
  \label{eq:CavityMeanField}
\vc{E}=\vc{D}/\epsilon^\prime=\overline{\vc{E}}.
\end{equation}
The resulting electric field inside the cavity can be computed in a
standard fashion:\cite{Jackson1975}
\begin{equation}\label{eq:onsager:F}
  \mathbf{F}=\frac{3\epsilon^\prime}{2\epsilon^\prime+\epsilon}
  \overline{\vc{E}}+\frac{2\lp\epsilon^\prime-\epsilon\rp} 
  {a^3\epsilon\lp2\epsilon^\prime+\epsilon\rp}  \mathbf{d},
\end{equation}
where the first term on the r.h.s. gives the field $\overline{\vc{E}}$
modified by the dielectric discontinuity at the cavity-solvent
interface while the second term is the image field of the dipole due
to polarization at the interface. The potential energy of the dipole
subsequently reads
\begin{equation}\label{eq:onsager:W}
  \pE=-\mathbf{d}\mathbf{F}=-\frac{3\epsilon^\prime}
  {2\epsilon^\prime+\epsilon}
  \mathbf{d}\overline{\mathbf{E}} - \frac{2 \lp \epsilon^\prime-\epsilon\rp}
  {a^3\epsilon\lp2\epsilon^\prime+\epsilon\rp}d^2.
\end{equation}
We can now calculate the average dipole moment and, via
Eq.~\eqref{eq:onsager:cavity:thepoint} and \eqref{eq:dipole:Chi}, the
susceptibility $\chi_{ij}$. In the high temperature
limit,\cite{Onsager1936}
\begin{equation}\label{eq:onsager:d} \langle \mathbf{d} \rangle =
  \frac{\int \dO \mathbf{d} e^{-\beta \pE}}{\int \dO e^{-\beta
      \pE}}\approx\frac{d^2
    \beta\epsilon^\prime}{2\epsilon^\prime+\epsilon}\overline{\mathbf{E}},
\end{equation} 
where the integration is over all possible orientations of
$\mathbf{d}$ and $\dO$ denotes an infinitesimal element of the
corresponding solid angle. Note that the image field in
Eq.~\eqref{eq:onsager:W} does not affect the orientation of the
dipole. We thus obtain for the susceptibility
\begin{equation}
\label{eq:dipole:derivative}
\begin{split}
  \chi_{ij}=\frac{\partial \overline{P}\,\!^\tsc{d}_i}{\partial
    \overline{E}_j}\Bigg\vert_{\overline{\vc{E}}=0}&=
  \frac{3c\beta\epsilon^\prime}{2\epsilon^\prime+\epsilon}\lb\langle
  d_id_j\rangle-\langle d_i\rangle\langle
  d_j\rangle\rb\Bigg\vert_{\overline{\vc{E}}=0} \\&= \frac{c \beta d^2
    \epsilon^\prime}{2\epsilon^\prime+\epsilon}\delta_{ij},
\end{split}
\end{equation}
which, upon substitution into Eq.~(\ref{eq:dipole:ee3}), produces the
following relation between the bare and full dielectric constants,
\begin{equation}\label{eq:onsager:simple:0}
  \epsilon^\prime=\epsilon+4\pi c\beta
  d^2\frac{\epsilon^\prime}{2\epsilon^\prime+\epsilon}.
\end{equation}
As a result,
\begin{equation}\label{eq:onsager:simple} \frac{\epsilon^\prime}{\epsilon} 
  =\frac{1}{4}\lp1+ \frac{b}{\epsilon}+ \sqrt{\lp1+\frac{b}{\epsilon}\rp^2+8}\rp,
\end{equation}
where 
\begin{equation} b \equiv 4\pi c\beta d^2.
\end{equation}
Equation~(\ref{eq:onsager:simple}) yields Eq.~(26) from Onsager's
paper,\cite{Onsager1936} if we neglect the polarizability and set
$\epsilon$ to unity.

In the above procedure, one integrates out local degrees of freedom to
determine the bulk response of the material. It is thus possible to
interpret the Onsager construction as a coarse-graining
procedure. From this viewpoint, one may regard relation
\eqref{eq:onsager:simple} as a {\em renormalization} of the dielectric
response due to local dipolar sources.  For infinitesimal values of
the parameter $b$, the renormalization flow looks particularly simple:
\begin{equation}
  \label{eq:el:renormalization}
  \epsilon^\prime=\epsilon+\frac{b}{3},
\end{equation}
Since $b$ is positive, the ``renormalization flow'' has a single,
``infinite-temperature plasma'' fixed point at
$\epsilon^\prime\to\infty$, where the Coulomb interaction is
completely screened. The physical reason for this up-renormalization
of the dielectric response is that molecular dipoles are directed, on
average, {\em along} the field thus screening the field locally.

\section{Theory of Elasticity: Analogy with Electrostatics and
  Digression on Notation and Tensor Algebra}
\label{sec:symm-betw-electr}

E.~Kr{\"o}ner\cite{kroner66} has pointed out analogies between
equations of electrostatics and continuum mechanics.  These analogies,
which are summarized in Table~\ref{tab:symm}, do not amount to a full
correspondence, nevertheless, which has to do with more than just the
difference in the tensor ranks of the objects in the two theories.
The most basic objects of electrostatics and continuum mechanics are
electric charge density $\rho$ and body force $\vc{f}$ respectively.
The former is the divergence of a vector, while the latter of a
tensor, viz.:
\begin{equation}
4\pi\rho = D_{i,i}
\end{equation}
and
\begin{equation}
  \label{eq:bal:eq:kroner}
  f_i = -\sC_{ij,j},
\end{equation}
where $\vc{D}$ is the dielectric displacement and $\sC_{ij}$ the
elastic stress tensor.

The material relation $\vc{D}=\epsilon\vc{E}$ in a dielectric
corresponds to Hooke's law in elasticity:
$\sC_{ij}=\tzC{C}_{ijkl}\eC_{kl}$, so that the tensor $\eC_{ij}$ plays
the role analogous to the electric field $\vc{E}$ while the rank-four
tensor of elastic moduli $\tzC{C}_{ijkl}$ is analogous to the
dielectric susceptibility, which is generally a 2nd rank tensor.  Here
we assume an isotropic dielectric medium for simplicity, so that the
dielectric susceptibility tensor is proportional to the unit matrix,
effectively allowing us to regard $\epsilon$ simply as a scalar. In
the case of isotropic elasticity, some simplification is {\em also}
possible, to be discussed shortly; still, the elastic response will
have to be written out explicitly as a rank-four tensor.

Of particular importance are expressions for the free energy; to write
these down we must choose an appropriate ensemble. For instance, in
electrostatics one may choose to work at fixed charge or fixed
field.~\cite{LLcont} The latter is more convenient in the present
context as we probe the response of the material to externally imposed
field. Likewise, it will be convenient to work at fixed deformation in
the elastic case. The resulting expression for the free energy
increments are:\cite{LLcont, LLelast}
\begin{equation} \label{dF} dF = - \frac{1}{4\pi} \vc{D}d\vc{E}
  \QQText{and} dF = +\sC_{ij} d\eC_{ij}.
\end{equation}
We have deliberately emphasized the distinct signs in front of the two
increments for they are ultimately responsible for the difference in
how the response functions are renormalized in the two descriptions in
the presence of non-removable dipole moments and sources of stress
respectively.  We shall see that the elastic deformation is {\em
  enhanced} by the presence of built-in stress, in contradistinction
with electrostatics.

\begin{table}
\begin{tabular}{|l|l|}
  \hline
  ~Electrostatics & ~Elasticity\\
  \hline
  ~~~$E_i$ & ~~~$\eC_{ij}$\\
  ~~~$D_i$ & ~~~$\sC_{ij}$\\
  ~~~$4\pi P_i$ & ~~~$\fsC_{ij}$\\
  ~~~$\rho$ & ~~~$f_i$\\
  ~~~$D_{i,i}=4\pi\rho$ & ~~~$\sC_{ij,j}=-f_i$\\
  ~~~$D_i=\epsilon{E_i}$ & ~~~$\sC_{ij}=\tzC{C}_{ijkl}\eC_{kl}$\\
  ~~~$\rho=-P_{i,i}$ & ~~~$f_i=\fsC_{ij,j}$\\
  ~~~$\rot\vc{E}=0$ & ~~~$\inc\e=0$\\
  ~~~$dF = - \frac{1}{4\pi}\vc{D}d\vc{E}$ & ~~~$dF = +\sC_{ij} d\eC_{ij}$ \vspace{0.5mm} \\ 
  \hline
\end{tabular}
\caption{\label{tab:symm} Analogy between  
  electrostatics and linear elasticity. Here $\vc{E}$ and 
  $\vc{D}$ are the electric field and displacement vectors 
  respectively. $\vc{P}$ 
  is the electric polarization, $\rho$  electric 
  charge density, $\eC_{ij}$ and $\sC_{ij}$ elastic strain and stress 
  tensors respectively, 
  $\vc{f}$ the body force, $\fsC_{ij}$ the internal stress, 
  $\epsilon$ the dielectric susceptibility, $\tzC{C}_{ijlm}$ 
  the  elastic moduli tensor, and $dF$  the free energy  increment.}
\end{table}

The internal, or ``built-in'' stress in glassy materials can be
introduced analogously to how polarization is introduced in the
electrodynamics of continuous media.\cite{LLcont} In charge neutral
dielectrics, $\int\dV\rho\ofR=0$, and, hence, $\rho$ must be the
divergence of a vector, $\rho=-P_{i,i}$, that vanishes outside the
dielectric.\cite{LLcont} Similarly, the volume average of the built-in
body force $\vc{f}(\vc{r})$ in glassy materials vanishes in the
absence of an external load, $\int\vc{f}\ofR\dV=0$, and, of course,
vanishes outside. Hence, the force $\vc{f}$ is also a divergence, but
of a tensor, which we call $\fsC_{ij,j}$:
\begin{equation}\label{eq:bodyforce}
  f_i=\fsC_{ij,j}.
\end{equation}
$\fsC_{ij}$ vanishes at the surface of a sample. The force on
particles resulting from the built-in stress must be exactly balanced
out by the restoring force of the lattice. By
Eq.~\eqref{eq:bal:eq:kroner},
\begin{equation}\label{eq:baleq}
  \sC_{ij,j}+f_i=\lp\sC_{ij}+\fsC_{ij}\rp_{,j}=0.
\end{equation}
Thus, $\fsC_{ij}$ is an internal (built-in) stress distribution
characterizing the state of an amorphous structure similar to how
polarization $\vc{P}$ characterizes the state of a dielectric
material. One can think of the deformation corresponding to the stress
tensor $\sC_{ij,j}$ as the response of the lattice to a defect in the
form of built-in stress. This deformation is compatible, in the sense
of Eq.~\eqref{eq:SaintVenant}, while the deformation corresponding to
the built-in stress itself is not.


To simplify notations, in the following we shall employ Walpole's
conventions.\cite{Walpole1981} In addition to using Einstein's
repeated index convention for tensor multiplication, we will drop
indexes in inner products altogether. For instance, we often write
$\s=\tz{AB}\e$ instead of
$\sC_{ij}=\tzC{A}_{ijkl}\tzC{B}_{klpq}\eC_{pq}$, and $\s\e$ instead of
$\sC_{ij}\eC_{ij}$. To avoid confusion we use upright fonts to label
tensors (both fourth- and second-rank) whenever their indexes are not
written out explicitly. Second-rank tensors are always denoted by bold
lower case Greek letters, while fourth-rank tensors are denoted by
capital sans-serif letters. As usual, the bold upright serif font is
reserved for 3-vectors. We use the corresponding italic fonts for
tensor and vector components. Note, that some of the Greek letters are
conventionally reserved for scalars. For instance, $\mM$ and $\Bm$
label shear and bulk moduli, $\nM$ is the Poisson ratio, and $\beta
\equiv ({\kB}T)^{-1}$. We also use index notations to label spatial
derivatives, e.g. the derivative $\partial\fsC_{ij}/\partial x_k$ of a
second-rank tensor field $\fs\ofR$ is denoted with $\fsC_{ij,k}$.

Most of the rank-4 tensors to be used below are isotropic. For these,
algebra can be greatly simplified in the following
way.\cite{Walpole1981} Consider for instance the elastic moduli
tensor $\cM$ of an isotropic medium with Lam\'{e} coefficients
$\lambda$ and $\mu$,
\begin{equation} \label{Liso}
\tzC{C}_{ijkl} = \lambda \delta_{ij}
\delta_{kl} + \mu \lp\delta_{ik} \delta_{jl} + \delta_{il}
\delta_{jk}\rp.
\end{equation}
Hereby the elastic energy density,\cite{LLelast}
\begin{equation} \label{felast} e = \half \eC_{ij} \tzC{C}_{ijkl}
  \eC_{kl},
\end{equation}
contains the only two scalars one can form using the entries of the
deformation tensor:\cite{LLelast}
\begin{equation} \label{Felast} e = \frac{\lambda}{2}\eC_{ii}^2 + \mu
  \eC_{ij}^2 = \frac{\kappa}{2}\eC_{ii}^2 + \mu (\eC_{ij} -
  \frac{1}{3} \delta_{ij} \eC_{kk})^2,
\end{equation}
where $\Bm$ is the bulk modulus:
\begin{equation}
  \label{eq:def:Bm}
    \Bm=\lambda+\frac{2}{3}\mu, 
\end{equation}
and $\mu$ is the shear modulus. The second equality in
Eq.~\eqref{Felast} is a convenient formulation of the free energy in
that the first term on the r.h.s. corresponds to pure uniform
compression/dilation while the second term to pure shear.
 
Only two of the non-zero entries of tensor \eqref{Liso} are
independent. It turns out that any isotropic tensor can be presented
as the following spectral decomposition:\cite{Walpole1981}
\begin{equation}
  \label{eq:def:L}
  \Lt{a}{b} = a \tz{J} + b \tz{K},    
\end{equation}
where $a$ and $b$ are some coefficients. The fourth-rank isotropic
tensors $\tz{J}$ and $\tz{K}$,
\begin{align*}
  \tzC{J}_{ijkl} & = \frac{1}{3}\delta_{ij}\delta_{kl},\\
  \tzC{K}_{ijkl} & = \half\lp
  \delta_{ik}\delta_{jl}+\delta_{il}\delta_{jk}-\frac{2}{3}\delta_{ij}\delta_{kl}\rp,
\end{align*}
are idempotent, i.e., they satisfy relations
\begin{equation}\tz{JJ}=\tz{J}\QQText{and}\tz{KK}=\tz{K},
\end{equation}
and mutually ``orthogonal'', 
\begin{equation}
\tz{JK}=\tz{KJ}=0.  
\end{equation}
Acting on a symmetric second-rank tensor, say $\bm{\upupsilon}$, the
tensors $\tz{J}$ and $\tz{K}$ extract its hydrostatic (diagonal) and
deviatoric (trace-less) parts respectively
\begin{equation}\label{eq:JK:action}
\begin{split}
  &\tzC{J}_{ijkl}\upsilon_{kl}=\frac{1}{3}\upsilon_{kk}\delta_{ij},\\
  &\tzC{K}_{ijkl}\upsilon_{kl}=
  \upsilon_{ij}-\frac{1}{3}\upsilon_{kk}\delta_{ij} \equiv
  {^\prime\!\upsilon_{ij}}.
\end{split}
\end{equation}
In this notation, free energy \eqref{Felast} looks particularly
simple:
\begin{equation} \label{Felast1} e = \frac{\kappa}{2}\eC_{ii}^2 + \mu
  \, {^\prime\!\eC_{ij}}^2.
\end{equation}
Consequently, the elastic moduli tensor \eqref{Liso} can be written as
\begin{equation}\label{eq:c:general}
  \cM=
  \Lt{3\Bm}{2\mu}=2\mu\;\Lt{\frac{1+\nu}{1-2\nu}}{1},
\end{equation}
where $\nu$ is the Poisson ratio of the medium:
\begin{equation}
  \label{eq:def:Pr}
  \nu \equiv \frac{1}{2\lp1+\mu/\lambda\rp}=\frac{3\Bm-2\mu}{2(3\Bm+\mu)}.
\end{equation}

Decomposition \rf{eq:def:L} simplifies the algebra for isotropic
tensors considerably. For example, for two isotropic tensors
$\tz{L}_1=\Lt{a_1}{b_1}$ and $\tz{L}_2=\Lt{a_2}{b_2}$, the
sum and the product are given simply by
\begin{equation}
  \tz{L}_1+\tz{L}_2=\Lt{a_1+a_2}{b_1+b_2}
\end{equation}
and 
\begin{equation}
  \tz{L}_1\tz{L}_2=\Lt{a_1a_2}{b_1b_2}
\end{equation}
respectively.  Also, the tensor equation $\tz{L}_1=\tz{L}_2$ is
equivalent to the system of two scalar equations, $a_1=a_2$ and
$b_1=b_2$. The $n$-th power of the tensor can be computed using the
formula
\begin{equation}\label{eq:L:power}
\Lt{a}{b}^{n}=\Lt{a^n}{b^n}, 
\end{equation}
where, note, $n$ can be non-integer.  Note that isotropic tensors
commute with each other, a property which also holds for the cubic
symmetry, but not so for other point symmetry
groups.\cite{Walpole1981}

\section{Renormalization of the elastic moduli by built-in stress:
  setting up a continuum description}
\label{sec:elasticity:general}

Let us now consider a degenerate equilibrium aperiodic medium. The
degeneracy is understood in the following way: The sample has a large
number of alternative ground states, all of which are minima of the
free energy. For each value of the free energy, there are an
exponential number of alternative aperiodic minima. Such situation is
realized in glassy liquids, where the number of alternative aperiodic
states for a sample of volume $V$ is given by $e^{\tilde{s}_c V/k_B}$,
where $\tilde{s}_c$ is the configurational entropy of the liquid per
unit volume. This entropy can be determined approximately by
calorimetry, see Refs.~\onlinecite{Capaccioli, RWLbarrier} and
references therein.  The aperiodic free energy minima are metastable
with respect to transitions between each other. These minima are also
metastable with respect to the crystalline state, if any; throughout,
we assume the nucleation barrier for crystallization is infinitely
high.

Because the individual aperiodic minima are metastable, they are
stable with respect to small, elastic deformation. The corresponding
elastic moduli tensor is denoted with $\cI$. We will refer to these
elastic constants as the ``bare'' constants. For simplicity, we assume
they do not vary between minima, i.e., the minima are distinct but
equivalent. In the language of replica-symmetry breaking, this
equivalence corresponds to one-stage replica symmetry
breaking.\cite{SpinGlassBeyond, MCT1, mezard:1076} Description at this
low-stage replica-symmetry breaking is believed to be adequate in
equilibrated liquids above the glass transition.\cite{MCT1, MCT, KTW,
  doi:10.1021/jp402235d, WileyBook}

\begin{figure}[t]
  \includegraphics[width=0.7\figurewidth]{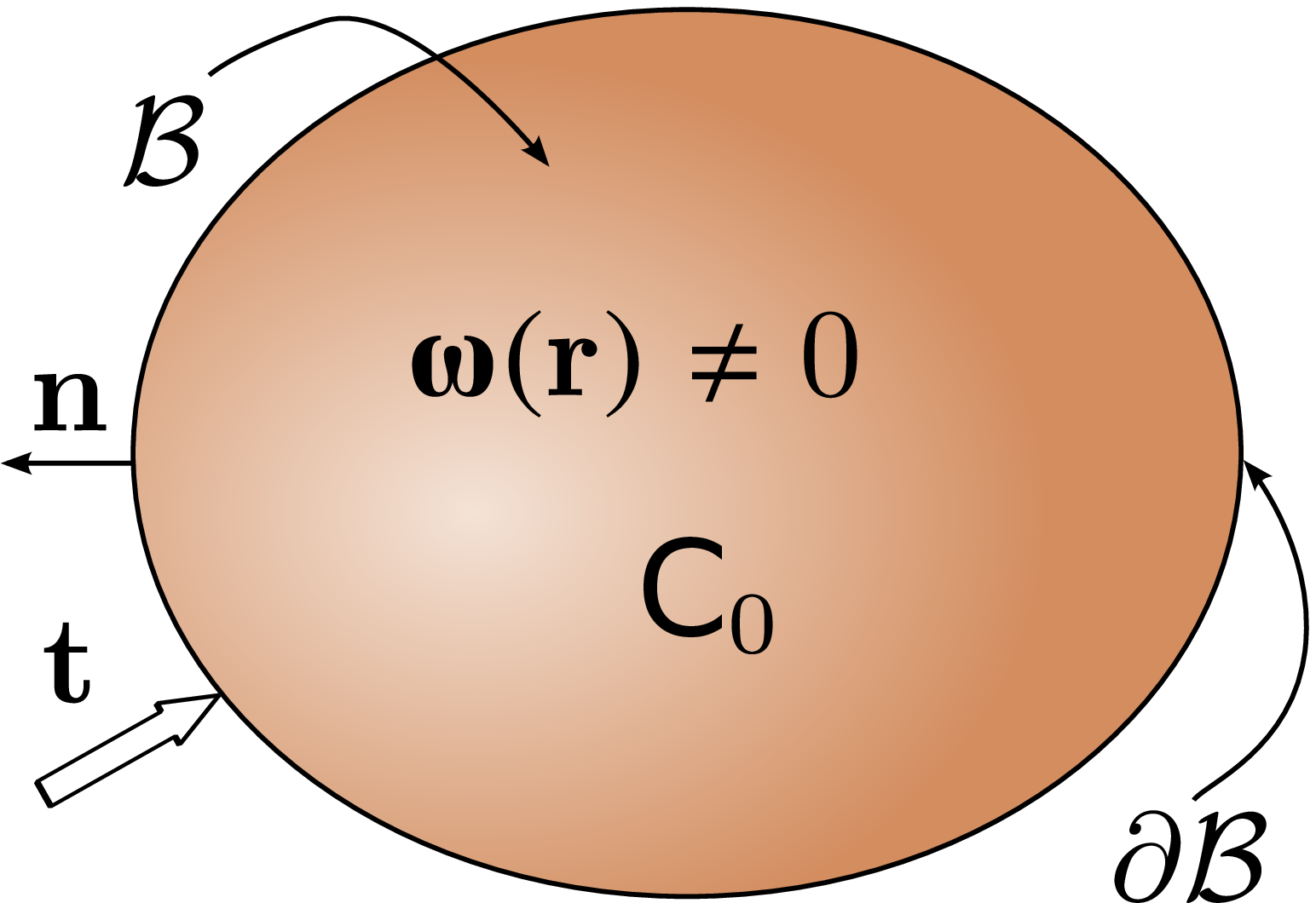}
  \caption{Setup of Section~\ref{sec:elasticity:general}: A
    homogeneous body $\Bo$ is subjected to external load in the form
    of a traction force $\vc{t}$ applied to the boundary $\dB$ of the
    body. The bare elastic constants of individual, aperiodic free
    energy minima are contained within the the fourth-rank tensor
    $\cI$.  Unit vector $\vc{n}$ is an external normal to $\dB$. The
    built-in stress, denoted by the tensor field $\fs\ofR$,
    corresponds to the mismatch penalty at spatial interfaces between
    distinct free energy minima.}
  \label{fig:modulidef}
\end{figure}

Consider a macroscopic sample $\Bo$ of an equilibrated aperiodic solid,
whose physical boundary is denoted with $\dB$. The internal---or
``built-in''---stress, due to spatial interfaces between distinct
aperiodic minima, is denoted with $\fs\ofR$, see
Fig.~\ref{fig:modulidef}. The RFOT theory has quantitatively
characterized the activated reconfigurations between the distinct
aperiodic minima, see review in Ref.~\onlinecite{LW_ARPC}. The presence of
the built-in stress modifies the elastic response of the body
analogously to how molecular dipoles modify the dielectric response of
the liquid.  Owing to the activated dynamics in the liquid, the
built-in stress pattern is not steady, but relaxes on the time scale
$\tau_\alpha$ of the $\alpha$-relaxation, even though the stress
magnitude is steady on average.  This is analogous to how polar
molecules can rotate in a solution.

If an external traction force $\mathbf{t(r)}$ is applied to the
boundary $\dB$ of $\Bo$, the resulting strain field $\mathbf{u}$ is a
solution of the boundary value problem,
\begin{equation}\label{eq:BVP}
\begin{dcases}
    \lp\sC_{ij}+\fsC_{ij}\rp_{,j}=0,\\
    \sC_{ij}n_j\Big\vert_\dB=t_i,
\end{dcases}
\end{equation}
supplemented by the constitutive relation (Hooke's law):
\begin{equation}
  \label{eq:sC0e}
\s=\cI\e.  
\end{equation}
Equation~\eqref{eq:BVP} is Newton's 3rd law, the top and the bottom
entry corresponding to the bulk and surface response respectively.
The unit vector $\mathbf{n}$ is an external normal to $\dB$. The
quantities $\s$ and $\e$ are, respectively, the elastic stress and
strain inside $\Bo$. The strain $\e$ is defined in Eq.~\eqref{eq:def}.

Since individual minima respond purely elastically, the strain field
$\eC_{ij}$ is \emph{compatible}, $\inc\e=0$,
cf. Eq.~\eqref{eq:SaintVenant}. In contrast, the strain $\fsn\ofR$
that corresponds to the internal stress $\fs\ofR$, $\fs=\cI\fsn$, can
not be represented as a derivative of a single-valued deformation
field. The field $\fsn$ is thus \emph{incompatible}:
\begin{equation}
  \label{eq:incompatibility2}
  \lp\inc\fsn\rp_{ij}=-\epsilon_{ikl}\epsilon_{jmn}\fsnC_{ln,km}\neq0.
\end{equation}
Both $\vc{t}\ofR$ and $\fs$ cause deformation in $\Bo$, as already
mentioned. Consequently, the elastic stress $\s$ is a sum of two
components:
\begin{equation}\label{eq:s}
\s=\sT+\sS,
\end{equation}
where the stress $\sT$, produced by the surface traction, obeys
\begin{equation}
  \label{eq:sT}
  \begin{dcases}
    \sTC_{ij,j}=0,\\
    \sTC_{ij}n_j\Big\vert_\dB=t_i,
  \end{dcases}
\end{equation}
while the stress $\sS$, produced by the source field $\fs$, satisfies
\begin{equation}
  \label{eq:sS}
  \begin{dcases}
    \lp\sSC_{ij}+\fsC_{ij}\rp_{,j}=0,\\
    \sSC_{ij}n_j\Big\vert_\dB=0.
  \end{dcases}
\end{equation}
The equation above follows from Eqs.~\eqref{eq:BVP} and
\eqref{eq:sT}. The elastic strain $\e$ can be similarly written as a
sum of two components,
\begin{equation}
  \label{eq:def:e}
  \e=\eT+\eS,
\end{equation}
where the strain produced by the traction force $\mathbf{t}$ and the
internal stress $\fs$ are defined as 
\begin{equation}
  \label{eq:def:sTandsS}
  \sT=\cI\eT\QQText{and}
  \sS=\cI\eS
\end{equation}
respectively. No built-in sources of stress lie at the boundary $\dB$
of the sample,
\begin{equation}\label{eq:fs:BC}
\fs\Big\vert_\dB=0,
\end{equation}
and the present analysis is limited to symmetric sources
\begin{equation}
  \label{eq:fs:symmetry}
 \fsC_{ij}=\fsC_{ji}. 
\end{equation}
The above boundary conditions for $\fs$ are standard in treatments of
defects in solids.\cite{Bacon1979} These conditions entail an
important relation between the volume averages of $\sS$ and $\fs$,
\begin{equation}\label{eq:vol:av}
\overline{\sS} = \cI\overline{\eS}
= -\overline{\fs},
\end{equation}
which is straightforward to show by writing
$\sSC_{ij}=\sSC_{ik}x_{k,j}$ and using Gauss's theorem together with
Eq.~\eqref{eq:sS}.\cite{Mura1987} Hereafter we use bars to indicate
averaging over the volume $V$ of $\Bo$. For instance,
\begin{equation}
  \label{eq:def:VolAv}
  \overline{\fs}=\frac{1}{V}\int_\Bo\fs\;\dV.
\end{equation}
Since our liquid is equilibrated, ensemble averaging is equivalent to
time averaging.

By Eq.~\eqref{eq:vol:av}, the built-in stress pattern $\overline{\fs}$
automatically reflects the symmetry of $\cI$ in the limiting cases of
a uniform liquid, $\mI=0$, and of an infinitely compressible body,
$\kI=0$. Indeed, $\mI=0 \Rightarrow \cI\propto\tz{J}$, and so
Eq.~\eqref{eq:JK:action} implies that for any $\overline{\eS}$, the
internal stress $\overline{\fs}$ is purely hydrostatic. Likewise, in
the other extreme $\kI=0 \Rightarrow \cI\propto\tz{K}$, the tensor
$\overline{\fs}$ is purely deviatoric.

Next, we determine the linear response of body $\Bo$ to an external
load.  We define the effective elastic moduli $\cM$ of $\Bo$ as a
fourth-rank tensor connecting the volume average of the \emph{total}
stress in $\Bo$ with the volume average of the total elastic strain:
\begin{equation}\label{eq:def:CM}
  \int_\Bo\lp \s+\fs\rp \dV = \int_\Bo\lp \cI\e+\fs\rp \dV \equiv \cM\;\int_\Bo \e\; \dV.
\end{equation} 
Note that by definition, $\cM$ is spatially uniform and
Eq.~\eqref{eq:def:CM} is the elastic analog of Eq.~\eqref{eq:av:ee2}.
Definition \eqref{eq:def:CM} is equivalent to the relation
\begin{equation}\label{eq:def:CMVas}
\int_\Bo\sT \dV = \cM\; \int_\Bo \e\; \dV,
\end{equation}
which is easy to show using Eqs.~\eqref{eq:s} and
\eqref{eq:vol:av}. The equation above relates quantities directly
accessible in experiment: the average total load $\overline{\sT}$
applied to $\Bo$ and the average resulted deformation $\overline{\e}$
of $\Bo$.  Using Eq.~\eqref{eq:def:e}, we can further rewrite
Eq.~\rf{eq:def:CMVas} as
\begin{equation}\label{eq:def:CM2}
\cI \overline{\eT} = \cM(\overline{\eT} + \overline{\eS}).
\end{equation}

In full correspondence with the above discussion of the symmetry of
the built-in stress $\overline{\fs}$, $\cM\propto\tz{J}$ for a uniform
liquid, $\mI=0$, while $\cM\propto\tz{K}$ for an infinitely
compressible solid, $\kI=0$. We note that both cases correspond to
\emph{fixed} points on the $\nI\mapsto\nM$ mapping, where $\nI$ and
$\nM$ are the bare and effective values of the Poisson ratio. Indeed,
by Eq.~\eqref{eq:c:general}, two isotropic fourth-rank tensors can be
proportional to each other only if their Poisson ratios are
equal. This notion will resurface in
Section~\ref{sec:cavity:softening}.

Since $\cM$ should not depend on the configuration of the load and the
shape of $\Bo$, we may conveniently assume a homogeneous $\eT$. Under
these circumstances, the 2nd equality in Eq.~\rf{eq:def:CM} yields:
\begin{equation}\label{eq:def:CM2b}
\cM\overline{\e}=\cI\overline{\e}+\overline{\fs}.
\end{equation}
c.f.  Eq.~\eqref{eq:av:ee2}.

Thus, $\overline{\fs}$ is a function of the average strain
$\overline{\e}$ in the material. Analogously to
Eq.~\eqref{eq:dipole:P:expansion2}, one has in the linear-response
regime:
\begin{equation}\label{eq:fs:expansion}  \overline{\fsC}_{ij} 
  = \susSC_{ijkl} \overline{\eC}_{kl},
\end{equation}
where we define the static susceptibility $\susSC_{ijkl}$ according
to:
\begin{equation} \label{eq:def:Susceptibility} \susSC_{ijkl} \equiv
  \frac{\partial\overline{\fsC}_{lm}}
  {\partial\overline{\eC}_{kl}}\Bigg\vert_{\overline{\e}=0},
\end{equation}
c.f. Eq.~\eqref{eq:dipole:Chi}. This results, together with
Eq.~\eqref{eq:def:CM2b}, in a linear response-type relation between
the effective and ``bare'' elastic moduli of $\Bo$:
\begin{equation}\label{eq:def:CM3}
\cM=\cI+\susS.
\end{equation}
Here we have used that $\overline{\fs}=0$ in the absence of external
load.  Equation~\eqref{eq:def:CM3} is the elastic analog of
Eq.~\eqref{eq:dipole:ee3}. It is valid for any symmetry of the tensor
$\cI$. The following analysis is limited to isotropic elasticity,
which is the simplest, yet most relevant case for amorphous
materials. The ``bare'' elastic moduli, comprising the tensor $\cI$,
will be labelled $\mI$, $\kI$, and $\nI$; these are the shear and bulk
modulus, and the Poisson ratio, respectively.  We expect $\cM$ to be
isotropic as well, since, by definition, amorphous materials are
isotropic in the long-wavelength limit. Therefore, the susceptibility
$\susS$ must be an isotropic tensor to satisfy
Eq.~\eqref{eq:def:CM3}. Consequently, the tensor
equation~\eqref{eq:def:CM3} is equivalent to two scalar equations, as
discussed Section~\ref{sec:symm-betw-electr}. The effective moduli
comprising $\cM$ will be labelled $\mM$, $\kM$, and $\nM$. By
Eq.~(\ref{eq:def:CM3}) they can be determined with the knowledge of the
response function (\ref{eq:def:Susceptibility}).

\section{Interaction between sources of built-in stress}
\label{sec:cavity:main}

As in the dielectric case, calculation of the susceptibility
\eqref{eq:def:Susceptibility} requires a specific microscopic model
for the dynamics of $\fs$. Here we explicitly obtain such a
microscopic model, which is the elastic analog of the dipole-dipole
interaction in electrostatics.

In an earlier publication,\cite{BL_6Spin} which will be referred to as
BL, we have put forth a minimal ansatz for the stress distribution in
equilibrated amorphous systems,\cite{BL_6Spin} as explained in the
Introduction, see Eq.~\eqref{eq:Anzatz}. BL have shown that the
dependence of the free energy $F$ on the magnitude of built-in stress
$g$ is concave at small and large values of $g$, but has a convex
portion at intermediate values of $g$. The low and high-$g$ states can
thus be interpreted as distinct phases separated by a nucleation
barrier. The high $g$ phase is aperiodic and vastly degenerate, the
degeneracy originating from the multitude of mutual configurations of
the degree of freedom $\fsn$, which has 5 independent components, in
view of the constraint \eqref{eq:op:local}.  Given sufficient steric
stabilization for aperiodic structures, the high $g$ phase can be made
metastable implying the built-in stress can be {\em self-consistently}
finite.

The interaction between the anharmonic degrees of freedom $\fsn$ is
determined by integrating out the long-wavelength
motions:\cite{BL_6Spin}
\begin{equation}
  \label{eq:Anzatz:Ham}
  \Ham_0 = \Ham_\text{SE}
  + a^6\sum_{m<n} \fsn(\vc{r}_n)\cI\tz{G}(\vc{r}_m -
  \vc{r}_n)\cI\fsn(\vc{r}_n),
\end{equation}
where the double sums are over all bead pairs.  The quantity
$\Ham_\text{SE}$ is the self-energy of the built-in stress in the
absence of external load, see below.  The coupling $\tz{G}$ is the
Fourier transform,
\begin{equation} \label{Mijkl} \tz{G}(\vc{r}) = -
  \int_{|\vc{k}|\le\frac{\pi}{a}} \frac{d^3(\vc{k})}{(2\pi)^3}
  \cos(\vc{k} \vc{r})\; \widetilde{\tz{G}}(\vc{k}),
\end{equation}
of the following tensor:
\begin{widetext}
  \begin{equation} \label{MijklFourier} \widetilde{\!\tzC{G}}_{ijml} =
    \frac{1}{\mI} \left( \delta_{im}\hat{k}_j \hat{k}_l +
      \delta_{il}\hat{k}_j \hat{k}_m+ \delta_{jm}\hat{k}_i \hat{k}_l+
      \delta_{jl}\hat{k}_i \hat{k}_m- \frac{\lI +\mI}{\lI+2\mI}
      \hat{k}_i\hat{k}_j\hat{k}_m\hat{k}_l \right),
\end{equation}
where $\hat{\vc{k}} \equiv \vc{k}/|\vc{k}|$. Recasting
Eq.~\eqref{eq:Anzatz:Ham} in terms of the internal stress
$\fs=\cI\fsn$ we obtain
\begin{equation}\label{eq:fen:general}
  \Ham_0=\half\int_\Bo\dV\fs\cI^{-1}\fs+\frac{1}{2}
  \int_\Bo\dV\int_\Bo\dV^\prime 
  \fsC_{ij}\ofR \,
  \tzC{G}_{ijlm}(\mathbf{r}-\mathbf{r}^\prime)\,
  \fsC_{lm}(\mathbf{r}^\prime),
\end{equation}
where we assume the built-in stress $\fs$ and deformation $\fsn$ are
related by Hooke's law: $\fs = \cI \fsn$.  Switching from discrete
summation over bead sites to spatial integration is done according to
the prescription $a^3\sum\to\int\dV$. The first integral on the
r.h.s. of Eq.~\eqref{eq:fen:general} corresponds to the self-energy
from Eq.~\eqref{eq:Anzatz:Ham}.

The expression for the coupling $\tz{G}$ between local sources of
stress $\fs$, Eq.~\eqref{Mijkl}, can be written out explicitly as:
\begin{equation}
  \label{eq:def:coupling}
  \begin{split}
    \tzC{G}_{ijml}(\mathbf{r}-\mathbf{r}^\prime)
    &=\frac{1}{64\pi^3\mI\lp1-\nI\rp}
    \lp\delta_{ip}\frac{\partial}{\partial x_j} +
    \delta_{jp}\frac{\partial}{\partial
      x_i}\rp\lp\delta_{ql}\frac{\partial}{\partial x_m}
    +\delta_{qm}\frac{\partial}{\partial x_l}\rp \\&\times\lp
    2\lp1-\nI\rp\delta_{pq}\frac{\partial^2}{\partial x_sx_s}-
    \frac{\partial^2}{\partial x_px_q}\rp
    \int_{\abs{\vc{k}}\le\frac{\pi}{a}}\rmd\mathbf{k}
    \frac{e^{i\mathbf{k}(\mathbf{r}-\mathbf{r}^\prime)}}{k^4}.
  \end{split}
\end{equation}
In the long-wavelength limit, $(r \rightarrow \infty) \Rightarrow (a
\rightarrow 0)$, the above expression simplifies significantly as the
integral reduces to $\int\rmd\mathbf{k}\exp\!\lb
i\mathbf{k}\mathbf{r}\rb/k^4=-\pi^2r$.\cite{Teodosiu1982} The
stress-stress coupling, which we label in this approximation by
$\tz{G}^\tsc{a}$, can be now expressed via the well known Green tensor
$\gA$ for a point force inside an infinite, homogeneous, and isotropic
medium (Kelvin's solution)\cite{Teodosiu1982} with the elastic moduli
tensor $\cI$:
\begin{equation} \label{eq:coupling:reg}
  \tzC{G}^\tsc{a}_{ijml}(\mathbf{r}-\mathbf{r}^\prime)=
  \frac{1}{4}\lp\delta_{ip}\frac{\partial}{\partial x_j} +
  \delta_{jp}\frac{\partial}{\partial
    x_i}\rp\lp\delta_{ql}\frac{\partial}{\partial x_m}
  +\delta_{qm}\frac{\partial}{\partial
    x_l}\rp\gAC_{pq}(\mathbf{r}-\mathbf{r}^\prime),
 \end{equation}
\end{widetext}
where 
\begin{equation}
  \label{eq:def:kelvin}
  \gAC_{ij}\ofR=\dfrac{1}{16\pi\mI\lp1-\nI\rp}
  \dfrac{1}{r}\lb\lp3-4\nI\rp\delta_{ij}
  +\dfrac{x_ix_j}{r^2}\rb.
\end{equation}
Note that, apart from the complicated tensorial form of the coupling,
the (long-wavelength) distance dependence of $\tz{G}^\tsc{a}$ is
$\propto 1/r^3$, analogously to the electric dipole-dipole
interaction. The tensor $\gA\ofR$ describes the response of the
elastic medium to point-force localized at the origin. Note that this
response diverges for a uniform liquid, $\mI=0$.  This is expected
since even an infinitesimal force causes an infinite displacement in a
uniform liquid, and so linear elasticity is no longer applicable. Note
that in the opposite extreme of an infinitely compressible body,
$\kI=0$ or $\nI=-1$, the kernel $\gA$ is well defined. As shown in
Appendix~\ref{sec:stress:source}, $\tz{G}^\tsc{a}\ofR$ is the tensor
describing the elastic response to a point-source of \emph{stress} by
an infinite, homogeneous, and isotropic medium. The elastic strain
$\eS$ produced by $\fs$ can thus be written as
\begin{equation} \label{eq:green:stress}
  \eS\ofR=\int_\Bo\dV^\prime\tz{G}^\tsc{a}
  (\mathbf{r}-\mathbf{r}^\prime)\fs(\mathbf{r}^\prime),
\end{equation}
so that the Hamiltonian \eqref{eq:fen:general} becomes
\begin{equation}
  \label{eq:fen:nocutoff}
  \Ham_0\approx \Ham^\tsc{a}_0=\half\int_\Bo\dV\fs\lp\eS+\cI^{-1}\fs\rp,
\end{equation}
where the superscript ``A'' indicates that the long-wavelength limit
of the Green's function is used.

In the presence of an external traction force, $\eT\neq0$, $\Ham_0$
must be supplemented by an appropriate coupling term.  We will show
systematically in Section~\ref{sec:cavity:gen}, see
Eq.~\eqref{eq:cav:en:homogen}, that this coupling is equal to the
expected $\int_\Bo\dV\eT\fs$.  Hence, the full Hamiltonian for the
internal stress in the presence of the external load reads
\begin{equation}
  \label{eq:ham:load}
  \Ham^\tsc{a}=\Ham^\tsc{a}_0+\int_\Bo\dV\eT\fs
  .
\end{equation}
The linear response of the system to externally imposed deformation
field $\eT\ofR$ is described, in the standard fashion, by the second
order isothermal response function
\begin{widetext}
  \begin{eqnarray} \label{eq:derivative:ham} \susC_{ijkl}(\vc{r},
    \vc{r}^\prime) &=& \frac{\delta\langle\fsC_{ij}\ofR\rangle}
    {\delta\eTC_{kl}(\vc{r}^\prime)}\Bigg\vert_{\eT=0}=
    \frac{\delta}{\delta\eTC_{kl}(\vc{r}^\prime)}
    \frac{\int\mathcal{D}\fs\,\fsC_{ij}\ofR\exp\lb-\beta
      \Ham^\tsc{a}\rb}{\int\mathcal{D}\fs\exp\lb-\beta
      \Ham^\tsc{a}\rb}\Bigg\vert_{\eT=0} \nonumber \\
    &=& -\beta \Big[ \langle \fsC_{ij}\ofR\fsC_{kl}(\vc{r}^\prime)
    \rangle_0 -\langle
    \fsC_{ij}\ofR\rangle_0\langle\fsC_{kl}(\vc{r}^\prime)
    \rangle_0\Big],
  \end{eqnarray}
\end{widetext}
where 
the naught on the r.h.s. indicates averaging in zero field, $\eT=0$,
by $\langle\cdots\rangle_0$. For instance,
\begin{equation}
  \label{eq:zero:field:average}
  \langle\fs\ofR\rangle_0=\frac{\int\mathcal{D}\fs\,\fs\ofR\,
    \exp\!\lb-\beta \Ham_0^\tsc{a}\rb}
  {\int\mathcal{D}\fs\,\exp\!\lb-\beta \Ham_0^\tsc{a}\rb}.
\end{equation}
By the chain rule of differentiation,
\begin{equation} \label{eq:fs:funexpansion}
  \langle\fsC_{ij}\ofR\rangle = \int \dV^\prime
  \frac{\delta\langle\fsC_{ij}\ofR\rangle}
  {\delta\eTC_{kl}(\vc{r}^\prime)}\Bigg\vert_{\eT=0}\eTC_{kl}(\vc{r}^\prime),
\end{equation}
and so the standard sum rule for the static isothermal susceptibility
$\widetilde{\sus}^\tsc{s}$ holds:
\begin{equation} \label{eq:derivative:hom0} \langle\fsC_{ij}\rangle =
  \eTC_{kl} \int\dV^\prime\susC_{ijkl}(\vc{r}, \vc{r}^\prime) =
  \widetilde{\!\susC}^\tsc{s}_{ijkl}\eTC_{kl}.
\end{equation}
We use the tilde to distinguish $\widetilde{\sus}^\tsc{s}$ from the
susceptibility defined by Eq.~\eqref{eq:def:Susceptibility} because it
corresponds to the derivative of $\langle{\fs}\rangle$ with respect to
the average field $\overline{\e}$, whereas $\widetilde{\sus}^\tsc{s}$
is equal to the derivative with respect to $\eT$. The two fields are
straightforwardly related by Eq.~\eqref{eq:def:CM2}.

\section{Cavity construction for supercooled liquids}
\label{sec:cavity:gen}

Evaluation of the cumulant in Eq.~\eqref{eq:derivative:ham} is
prohibitively difficult to accomplish analytically for the model in
Eq.~\eqref{eq:fen:general}.  However, because of the Coulomb-like
distance dependence of the interaction between point sources of force,
we may proceed by analogy with the electric dipole-dipole interaction,
for which Onsager's cavity method can be employed.  The analogy
between elasticity of stressed continua and electrostatics of polar
dielectrics was noticed a long time ago and used primarily to study
crystalline materials with a low concentration of
defects.\cite{kroner66,Kanzig1962} Periodic lattices are anisotropic,
which usually implies there at most few, discrete states of an
individual defect.\cite{Nowick1963} The same thing can be said about
orientational glasses, which are periodic crystals containing
anisotropic substitutional impurities, whose orientations are random
given sufficiently high density and/or low
temperature.\cite{LoidlARPC} Reorientational dynamics of such
impurities lead to a marked temperature dependence of the elastic
moduli.\cite{Lynden-Bell1994,LoidlARPC}

We have shown earlier,\cite{BL_6Spin} see also
Sec.~\ref{sec:cavity:softening} below, that constraint
\eqref{eq:op:local} is equivalent to fixing the length of a certain
$6$-component vector. Thus the structural dynamics of the built-in
stress correspond to the rotations of interacting $6$-vectors. This
makes an effective field approximation for the elasticity of
supercooled liquids conceptually very similar to that for the
dielectric properties of polar liquids. In fact, as we will show
below, it is possible to generalize Onsager's cavity construction to
find a relation between the bare and renormalized elastic moduli of a
supercooled liquid. This relation is controlled by the magnitude of
the built-in stress, similarly to how the bare and renormalized
dielectric constants are related via the magnitude of the molecular
dipole in Eq.~\eqref{eq:onsager:simple}.

\begin{figure}
  \includegraphics[width=\figurewidth]{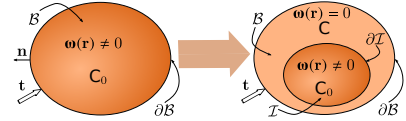}
  \caption {Generalized cavity construction for the elasticity of
    supercooled liquids. The original amorphous body $\Bo$,
    characterized by the bare elastic moduli tensor $\cI$ and
    containing spatially-distributed built-in stress $\fs$ is shown on
    the left. In the cavity construction, illustrated on the right,
    one treats explicitly a compact subset of the built-in sources of
    stress, while the response of the environment is approximated by
    that of an elastic continuum with effective constants $\cM$. The
    latter are found self-consistently for a given magnitude of the
    built-in stress.}
  \label{fig:effective:approx}
\end{figure}

Consider an ellipsoidal region $\I$ inside the body $\Bo$. We will
treat this region explicitly, whereby the region is characterized by
bare elastic constants $\cI$ and an intrinsic distribution of built-in
stress $\fs$. In contrast, the built-in stress {\em outside} the
region will be treated effectively; we will approximate the response
of the environment as an elastic continuum with effective elastic
constants $\cM$, see Fig.~\ref{fig:effective:approx}. Our aim is to
determine these effective elastic constants $\cM$ self-consistently.
We emphasize that this approach is not merely phenomenological, while
being surely consonant with observation. Its validity has the same
origin as the theory of dielectrics and stems from the fact that the
electric field due to an infinite, uniformly charged plane is
coordinate independent (within the individual half-spaces), leading to
a uniform polarization-induced field inside a polarizable slab subject
to a uniform external field. In turn, this notion stems from the $1/r$
dependence of the Green's function for electrostatic and elastic
interactions, which has to do with the lack of mass for photons and
phonons respectively.

Since both $\cI$ and $\cM$ are assumed to be isotropic, all possible
orientations of $\I$ in $\Bo$ are equivalent. The linear size of
$\Bo$, $L_\Bo$, is assumed to be much larger than that of $\I$;
consequently, we neglect the image forces produced by $\dB$ since the
corresponding contribution is $\propto L_\Bo^{-3}$.  A key feature of
the construction is that it neglects correlations between the sources
of built-in stress inside and outside of $\I$ for the purpose of
estimating $\la\fs\ra$.  This is a good approximation, if the size of
$\I$ is larger than the correlation length of the stress distribution
$\fs$. The thermodynamic average $\la\fs\ra$ is then approximated by
Boltzmann averaging over all structural states inside $\I$ only.

First off, the full elastic energy of $\Bo$ that contains an inclusion
$\I$ containing sources of built-in stress $\fs$ is equal to
\begin{equation}
  \label{eq:cavity:elastic}
  \pE_{el}=\frac{1}{2}\int_\Bo\dV\lp \s+\fs\rp\lp\e+\cI^{-1}\fs\rp,
\end{equation}
where the integrand is the product of the total (compatible and
incompatible) stress, $\s+\fs$, and the total strain,
$\e+\cI^{-1}\fs$.  The elastic stress $\s$ and strain $\e$ are
solutions of Eq.~\eqref{eq:BVP}, but since $\Bo$ now contains an
inhomogeneity in the form of the elastic discontinuity at the region
boundary $\prtl \I$, Eq.~\eqref{eq:s} no longer holds. The total
elastic stress now has to include a contribution from the stress
``polarization'' at the region boundary:
\begin{equation}
  \label{eq:cavity:s}
  \s=\sT+\sP+\sS,
\end{equation}
where 
\begin{equation}
  \label{eq:cavity:sT}
\sT=\cM\,\eT
\end{equation}
everywhere in $\Bo$; $\sT$ satisfies~\eqref{eq:sT}. The quantity
\begin{equation}
  \label{eq:sP}\sP=
  \begin{dcases}
    \cM\eP,\text{ outside }\I,\\
    \cI\eP,\text{  inside }\I,
  \end{dcases}
\end{equation}
is the stress produced by the boundary $\prtl \I$, while $\sS$ is the
stress produced by $\fs$.  As before, $\sS$ is given by the solution
of \eqref{eq:sS} but the constitutive relations are now different
between the region and the environment:
\begin{equation}
  \label{eq:cavity:sS}\sS=
  \begin{dcases}
    \cM\eS,\text{ outside }\I,\\
    \cI\eS,\text{  inside }\I.
  \end{dcases}
\end{equation}
Analogously to $\sS$, the stress $\sP$ must satisfy the free traction
boundary conditions on the surface of $\Bo$:
\begin{equation}
  \label{eq:BC:sP}
 \sPC_{ij}n_j\Big\vert_\dB=0.
\end{equation}

Next we use the cavity construction to evaluate the response function
\eqref{eq:def:Susceptibility}. First we need to establish a
correspondence between the homogeneous set-up of
Section~\ref{sec:elasticity:general} and the present situation with an
elastic discontinuity at the region boundary $\prtl \I$. The traction
forces in both cases are equal to each other analogously to how the
dielectric displacement is not modified, on average, by introducing a
cavity. Further, by Eqs.~\eqref{eq:cavity:sT} and
\eqref{eq:def:CMVas}, we establish that the traction displacement
$\eT$ outside the inclusion $\I$ corresponds with the average strain
$\overline{\e}$ defined in Section~\ref{sec:elasticity:general}. Thus,
the static susceptibility $\susS$ from
Eq.~\eqref{eq:def:Susceptibility} must be evaluated via
\begin{equation}
  \label{eq:cavity:susceptibility}
\susSC_{ijkl}=\frac{\partial\overline{\fsC}_{ij}}{\partial\eTC_{kl}}\Bigg\vert_{\eT=0}. 
\end{equation}

From here on, we assume $\sT$ and $\eT$ are homogeneous. Then, for an
ellipsoidal region $\I$, $\eP$ \emph{inside} $\I$ is also homogeneous
and is given by\cite{Eshelby1951, Walpole1981, Mura1987}
\begin{equation}
  \label{eq:eshelby}
  \eP=\tz{S}\tz{Q}\lp\cM-\cI\rp\eT,
\end{equation}
where 
\begin{equation} \label{eq:def:Q}
  \tz{Q} \equiv \lp \cM - \lb \cM - \cI \rb \tz{S} \rp^{-1},
\end{equation}
and $\tz{S}$ is the so-called Eshelby tensor.\cite{Walpole1981} The
Eshelby tensor appears in continuum mechanics as the solution to the
following problem: Imagine that a region $\I$ inside a homogeneous
elastic continuum with moduli $\cM$ experiences a structural
transformation, such as a martensitic transition. Under these
circumstances the region would relax to attain a uniform stress-free
strain $\e^\ast$, {\em if} removed from the matrix. What is the
deformation $\tilde{\e}$ of the region $\I$, if it remains inside the
matrix?  Eshelby has shown that\cite{Eshelby1957}
\begin{equation} \label{eq:def:eshelby:pure0} \tilde{\e}=\tz{S}
  \,\e^\ast,
\end{equation}
where the tensor $\tz{S}$ is generally a function of the coordinate
and depends on the shape of the region $\I$.  If $\I$ is an ellipsoid,
however, the Eshelby tensor is spatially uniform, and so is
$\tilde{\e}$. Despite its uniformity (for ellipsoidal $\I$), $\tz{S}$
is generally not isotropic, and so the order of multiplication in
Eq.~\eqref{eq:eshelby} matters. If, however, $\I$ {\em is} spherical,
$\tz{S}$ does become isotropic:
\begin{equation}\label{eq:S:sphere}
\tz{S}=\frac{1}{3\lp 1-\nM \rp}\;\Lt{1+\nM}{\frac{2}{5}\lp4-5\nM\rp}.
\end{equation} 
The last equation applies also when the elastic moduli experience a
discontinuity at the region boundary, the case we are interested here.
Note that $\tz{S}$ depends only on the Poisson ratio of the matrix,
i.e. the part of $\Bo$ outside $\I$.  Also, for $\nM=1/5$, the Eshelby
tensor is proportional to the unit tensor $\tz{I}$,
\begin{equation}
  \label{eq:EshTens:fifth}
  \tz{S}\Big\vert_{\nM=1/5}=\half\tz{I}, 
\end{equation} 
so that $\tilde{\e}=\e^\ast/2$, i.e., $\tilde{\e}$ and $\e^\ast$ are
related via a scalar. This is a peculiar situation, in which the
self-consistent tensor equation~\eqref{eq:def:CM3} boils down to a
single scalar equation, as we shall see in
Sec.~\ref{sec:cavity:softening}. Hereby the bulk and shear modulus are
renormalized in equal measure so that $\nM$ remains equal to
$\nI=1/5$.

In general, see Appendix~\ref{app:eshelby}, the Eshelby tensor is
related to the average of the Green tensor over the volume of the
inclusion,
\begin{equation}
  \label{eq:def:EshelbyTz}
\tz{S}=-\int_\I\dV^\prime\tz{G}^\tsc{a}(\mathbf{r}-\mathbf{r}^\prime)\cM, 
\end{equation}
where $\tz{G}^\tsc{a}$ is defined by Eq.~\eqref{eq:coupling:reg} with
$\mI$ replaced by $\mM$ and $\nI$ replaced by $\nM$. Note, that since
$\tz{G}^\tsc{a}\propto\mM^{-1}$, while $\cM\propto\mM$, see
Eqs.~\eqref{eq:coupling:reg}, \eqref{eq:def:kelvin}, and
\eqref{eq:c:general}, the tensor $\tz{S}$ depends only on the Poisson
ratio of the matrix for any shape of the region $\I$.

Below we will consider exclusively a spherical $\I$, in which case the
tensor $\tz{Q}$ becomes
\begin{equation}
  \label{eq:Q:ForSphere}
\begin{split}
  \tz{Q} = \frac{3 (1-\nM)}{2\mI}\;
  \tz{L}\Big[\;&\frac{1-2\nI}{(1+\nM)(1+\nI+2\smr\lb1-2\nI\rb)}\;,\\
  &\frac{5}{2\lb4-5\nM\rb+\smr\lb7-5\nM\rb}\;\Big].
\end{split}
\end{equation}

To determine the free energy $\pE$ proper of the built-in stress
pattern $\fs$ in the presence of external load $\mathbf{t}$ we need to
subtract from the full free energy $\pE_{el}$ in
Eq.~\eqref{eq:cavity:elastic} the elastic free energy of the body if
it were homogeneous: $(1/2)\int_\Bo \sT\eT\dV$, and the work $\int_\dB
\mathbf{t}\lp \mathbf{u}^\tsc{p}+\mathbf{u}^\tsc{s}\rp \dS$ of the
built-in stress and the stress due to the elastic discontinuity
expended to distort the boundary of the macroscopic body $\Bo$:
\begin{equation}
  \label{eq:cavity:pE}
  \pE=\pE_{el}- \frac{1}{2}\int_\Bo \sT\eT\dV - \int_\dB \mathbf{t}\lp
  \mathbf{u}^\tsc{p}+\mathbf{u}^\tsc{s}\rp \dS.
\end{equation}
Multiple application of Gauss's theorem together with
Eqs.~\eqref{eq:sT}-\eqref{eq:fs:BC}, \eqref{eq:BC:sP}, and
\eqref{eq:eshelby} allows one to recast $\pE$ in terms of an integral
over the inclusion only:
\begin{equation}\label{eq:EpGeneral}
\begin{split}
\pE = \half \int_\I \dV 
\Big( 
&\fs\eS+\fs\;\cI^{-1}\fs + 
\eT \cM\lp\cI-\cM\rp \tz{Q} \eT \\+ 
&\eT \lp \tz{I} + \cM\; \tz{Q}\rp \fs + \eT \lp\cI - \cM \rp \eS
\Big).
\end{split}
\end{equation}
The derivation of this equation can be found in Chapter 4 of Mura's
monograph;\cite{Mura1987} here we only briefly review the result
itself.  The first two terms in the integrand have the same form as
the earlier discussed Hamiltonian $\Ham^\tsc{a}_0$,
Eq.~\eqref{eq:fen:nocutoff},
\begin{equation}
  \label{eq:def:E0}
  \pE_0=\half\int_\I\dV\fs\lp\eS+\cI^{-1}\fs\rp.
\end{equation}
In important distinction from Eq.~\eqref{eq:fen:nocutoff}, the
integration is over the inclusion $\I$ only. The third term, $(1/2)\eT
\cM\lp\cI-\cM\rp \tz{Q} \eT$, is the potential energy due to the
elastic inhomogeneity in the absence of built-in stress, i.e., when
$\fs=0$; this term naturally vanishes for $\cI=\cM$.  Its sign is
determined by the relative values of the bare and renormalized elastic
constants $\cI-\cM$.  For instance, suppose that $\mI<\mM$, $\kI<\kM$,
and there is no built-in stress other than the elastic discontinuity,
i.e., $\fs=0$. Under these circumstances, the potential energy $\pE$
is {\em negative} signifying that introduction of the inhomogeneity
$\I$ makes the system (locally) unstable and may result, for instance,
in cracking. Apropos, the third term in Eq.~\eqref{eq:EpGeneral}
provides the basis for the Griffith fracture criterion for a spherical
inhomogeneity,\cite{Mura1987} whereby the growth of the crack is
limited by its surface energy.  The last two terms in the integrand in
Eq.~\eqref{eq:EpGeneral} 
describe the interaction between the built-in stress $\fs$ and the
externally imposed strain $\eT$.

Note that for a homogeneous $\Bo$, i.e. when $\cM=\cI$,
\begin{equation}\label{eq:cav:en:homogen}
  \pE\Big\vert_{\cM=\cI}=\pE_0+\int_\I\dV\eT\fs,
\end{equation}
since $\tz{I}+\cM\tz{Q}=2\tz{I}$ for $\cM=\cI$. This equation has the
same form as Eq.~\eqref{eq:ham:load}, showing that $\Ham^\tsc{a}$ is
indeed the correct Hamiltonian for the stress distribution subject to
an external field. Again, the integration in \eqref{eq:cav:en:homogen}
is over the volume of the inclusion $\I$, not the whole body $\Bo$.

We finish this Section by writing down a formal expression for the
susceptibility $\sus(\vc{r},\vc{r}^\prime)$ for the generalized cavity
construction shown on the right in Fig.~\ref{fig:effective:approx}. By
Eqs.~\eqref{eq:cavity:susceptibility} and \eqref{eq:EpGeneral}, we
obtain:
\begin{equation} \label{eq:derivative:cavity}
\begin{split}
  \susC_{ijkl}(\vc{r},\vc{r}^\prime)&=
  \frac{\delta}{\delta\eTC_{kl}(\vc{r}^\prime)}
  \frac{\int\mathcal{D}\fs\,\fsC_{ij}\ofR\exp\lb-\beta
    \pE\rb}{\int\mathcal{D}\fs\exp\lb-\beta \pE\rb}\Bigg\vert_{\eT=0}\\
  &= -\beta \Big[ \langle \fsC_{ij}\ofR\zeta_{kl}(\vc{r}^\prime)
  \rangle_0 -\langle
  \fsC_{ij}\ofR\rangle_0\langle\zeta_{kl}(\vc{r}^\prime)
  \rangle_0\Big],
\end{split}
\end{equation}
where 
\begin{equation}
  \label{eq:dif:E:zeta}
  \bm{\upzeta}=\half 
  \lb\lp\tz{I}+\cM\tz{Q}\rp\fs-\lp\cM-\cI\rp\eS\rb, 
\end{equation}
and $\langle\cdots\rangle_0$, again, denotes thermodynamic averaging
in zero external field, $\eT=0$.

The cavity construction is a reasonable approximation so long as the
correlation length for the stress-stress interaction does not exceed
the cavity size.  It is possible to systematically improve on this
approximation by including more sources of stress in the cavity $\I$
to explicitly account for many-particle effects. Such an approach can
be implemented in simulations, as has been noted in the context of
polar liquids.\cite{Friedman1975}

\section{Specific realization of the cavity construction: The case of
  uniform internal stress}
\label{sec:cavity:softening}

One typically visualizes the Onsager cavity construction as an
electric dipole in the center of an empty spherical cavity within a
continuum dielectric. However, one might equally well think of the
dipole moment due to the molecular dipole as uniformly {\em
  distributed} over the cavity.  This will only modify the image field
contribution to the energy of the molecular dipole. The image field is
however aligned with the dipole itself and does not affect its
orientation; Eq.~\eqref{eq:onsager:simple} thus still applies.  It
will be convenient to pursue this ``smeared source'' approach in the
elastic case as it readily produces closed form relationships between
the bare and renormalized elastic constants.  For a uniform $\fs$ and
a spherical inclusion $\I$,\cite{Eshelby1957,Walpole1981,Mura1987}
\begin{equation}\label{eq:eS:homogeneous}
\eS = - \tz{Q S} \fs,
\end{equation}
and so the energy of the built-in stress inside the cavity, from
Eq.~\eqref{eq:EpGeneral}, now reads:
\begin{equation}\label{eq:EpConst} \frac{\pE}{v} =
  \frac{1}{2}\fs\;\tz{G}\;\fs + \eT\;\cM\;\tz{Q}\;\fs
  +\frac{1}{2}\eT\cM\lp\cI-\cM\rp \tz{Q}\; \eT,
\end{equation}
where $v$ is the volume of $\I$, the tensor $\tz{G}$ is defined as
\begin{equation}\label{eq:def:G}
  \tz{G} \equiv \cM\lp \tz{I}-\tz{S} \rp \tz{Q\;} \cI^{-1},
\end{equation}
and the Eshelby tensor $\tz{S}$ is given
by~\eqref{eq:S:sphere}. Equation \eqref{eq:EpConst} is Eq.~(25.24)
from Mura's monograph\cite{Mura1987} written out explicitly for an
isotropic solid. We note that the first and second quadratic forms on
the r.h.s. of Eq.~\eqref{eq:EpConst} are positive definite, while the
third generally is not, as remarked earlier. Nevertheless, we shall see
this term is always positive for the renormalized values of $\mM$ and
$\kM$ that will be obtained self-consistently in the following.

The present, effective approach to the elasticity of aperiodic solids
is, of course, an approximation. The choice of detailed implementation
of the built-in stress is not unique and must be made depending on the
circumstances. 

In the first approach, we explicitly consider only a single source of
built-in stress that is in direct contact with the effective elastic
medium, analogously to the Onsager cavity construction. Hereby we fix
the magnitude of $\pE$ in the absence of external loading, $\eT=0$,
while assuming the elastic constants are $\cI$ and $\cM$ inside and
outside the cavity, respectively:
\begin{equation}\label{eq:ConstG} \pE^\tsc{g}=\frac{v}{2}\fs\tz{G}\fs
  \equiv \frac{\theta^\tsc{g}}{2 \beta} =\text{Const}.
\end{equation}

Despite similarities between the dielectric and elastic cases, there
is a fundamental distinction between how one can implement constraints
on local sources of built-in stress in the two descriptions.  In
contrast with the dielectric case, the stress energy in
Eq.~\eqref{eq:ConstG} also includes the deformation energy of the {\em
  environment}. Indeed, while the dipole moment of a standalone
molecule can be defined, the built-in stress within a small group of
molecules only if it is inserted in an ill-fitting elastic matrix; the
built-in stress thus cannot be defined on its own, i.e., without an
environment.

In the second approach, we also fix the magnitude of the self-energy
of the built-in stress in the absence of external load, but this time
we use the {\em bare} elastic constants both inside and outside of the
cavity. Substituting $\cM=\cI$ into the matrix $\tz{G}$ thus yields
the following constraint:
\begin{equation}\label{eq:ConstGlobal}
  \pEF \equiv \frac{v}{2} \fs\; \cI^{-1}\lp \tz{I}-\tz{S}_0
  \rp\fs  \equiv \frac{\theta^\tsc{f}}{2 \beta} =\text{Const}.
\end{equation} 
where $\tz{S}_0$ is the Eshelby tensor for a medium with elastic
moduli $\cI$. This way of constraining the built-in stress is
appropriate when we wish to consider more than one sources explicitly.
Clearly, each of these sources is inserted in the original medium
characterized by the bare constants.

The third type of the constraint is equivalent to the constraint from
Eq.~\eqref{eq:op:local} which corresponds to the original BL
model.\cite{BL_6Spin} Here one assumes that the self-interaction part
of $\pEF$ is fixed:
\begin{equation} \label{eq:constraint:local} \pES \equiv \frac{v}{2}
  \fs\cI^{-1}\fs=\frac{v}{2} \fsn\cI\fsn \equiv
  \frac{\theta^\tsc{s}}{2 \beta} =\text{Const},
\end{equation}
where $\fsn=\cI^{-1}\fs$, as discussed above, see
Eq.~\eqref{eq:op:local}.\cite{BL_6Spin} According to the preceding
discussion, this type of constraint does not explicitly consider the
contribution of the medium to the full cost of the built-in
stress. This may still be reasonable, if the inclusion size is large
enough to sustain built-in stress on its own.  Think of it as the
smallest size of a standalone molecular cluster that has long-lived
aperiodic minima, in addition to the lowest energy, crystalline
minimum. Despite its limitations, the ansatz from
Eq.~\eqref{eq:constraint:local} is of some formal value as it will
allow us to recast the minimalistic BL model in an actual continuum
fashion.

As already remarked in Section~\ref{sec:elasticity:general}, we assume
that all distinct aperiodic free energy minima are equivalent,
implying that we can take the values of the constants in
Eqs.~\eqref{eq:ConstG}-\eqref{eq:constraint:local} to be uniform in
space.  Below, we work out all three constraint types. The
calculations are straightforward but tedious; they are mostly
relegated to Appendix~\ref{app:Z:response}. The technical gist of the
calculation is as follows: We compute the partition function
corresponding to the energy from Eq.~\eqref{eq:EpConst}, which
requires integration with respect to the six components of the $\fs$
tensor, subject to the constraints from
Eqs.~\eqref{eq:ConstG}-\eqref{eq:constraint:local} for the three cases
respectively. In practice, this is best done by switching to a special
notation, in which rank-two tensors, such as $\fs$ and $\eT$, are
represented as six-component vectors, while the rank-4 tensors, such
as $\tz{G}$, become 6-by-6 matrices; the latter happen to transform as
tensors in this special notation.\cite{Voigt} The constraints
\eqref{eq:ConstG}-\eqref{eq:constraint:local} then amount to fixing
the magnitude of bilinear forms for the components of 6-vectors. This
constraint can be further simplified by a coordinate transformation in
the 6-space, upon which the quadratic form becomes the unit
matrix. Consequently, each constraint is equivalent to fixing the
length of a certain 6-vector, whose precise identity varies between
the three cases.

\subsection{Constraint 1: The ``Onsager'' limit}
\label{sec:constG}

The self-energy energy $\pEG$ from Eq.~\eqref{eq:ConstG} reads
explicitly, in terms of the components of the built-in stress $\fs$,
as
\begin{equation}\label{eq:Ep0:G}
\begin{split}
  \frac{\pEG}{v}&=\frac{2\mM}{9\kI} \, \frac{\fsC_{ii}^2}{3\kI+4\mM} \\
  &+\frac{\mM}{4\mI} \:
  \frac{9\kM+8\mM}{\mM\lp9\kM+8\mM\rp+6\mI\lp\kM+2\mM\rp} \:
  {}^\prime\!\fsC_{ij}^2,
\end{split}
\end{equation}
where 
\begin{equation}\label{eq:fs:deviatoric}
  ^\prime\!\fsC_{ij} \equiv \fsC_{ij}-\frac{1}{3}\fsC_{kk}\delta_{ij},
\end{equation}
is the deviatoric (trace-less) part of $\fs$.  
 
Clearly, in the limiting case of $\mI=0$ ($\nI=1/2$), only the
hydrostatic component of $\fs$ can be non-zero, consistent with the
physical expectation that a uniform liquid cannot sustain built-in
stress.  Likewise, in the case of an infinitely compressible liquid,
$\kI/\mI=0$ ($\nI=-1$), only the deviatoric part $\fsC_{ij}=
{^\prime\!\fsC_{ij}}$ is non-vanishing.  In the spirit of the
equipartition theorem, the two terms on the r.h.s. of
Eq.~\eqref{eq:Ep0:G} are expected to have comparable magnitudes.  We
thus tentatively conclude that as the Poisson ratio of the material
changes from $-1$ to $1/2$---which corresponds to a decreasing shear
modulus relative to the bulk modulus---the frozen-in stress pattern in
the form of the built-in stress $\fs$ switches character from mostly
frozen-in shear to mostly frozen-in uniform compression/dilation,
consistent with the BL analysis.\cite{BL_6Spin}

We show in Appendix~\ref{app:Z:response} that Eq.~\eqref{eq:def:CM3},
which connects the effective and bare elastic moduli, now becomes:
\begin{equation}\label{eq:constG:Teq}
\cM=\cI\lp\tz{I}-\frac{\theta^\tsc{g}}{6}\lb\tz{I}-\tz{S}\rb^{-1}\rp,
\end{equation}
where the Eshelby matrix for the effective medium is given in
Eq.~(\ref{eq:S:sphere}); it depends on the effective Poisson ratio
$\nM$. The above equation thus can be used to determine the effective
moduli self-consistently.

According to Sec.~\ref{sec:symm-betw-electr}, the tensor equation
\eqref{eq:constG:Teq} is equivalent to the system of two scalar
equations, viz.,
\begin{equation}\label{eq:constG:Aeqs}
\begin{dcases}
  \frac{\mM}{\mI}&=1+\theta^\tsc{g}\lp\frac{1}{7-5\nM}-\half\rp; \\
  \frac{\mM}{\mI}&=\frac{\lp1+\nI\rp\lp4-\theta^\tsc{g}
    \lb1-\nM\rb-8\nM\rp}{4\lp1-2\nI\rp\lp1+\nM\rp},
\end{dcases}
\end{equation}
where the dimensionless magnitude $\theta^\tsc{g}$ of the built-in
stress energy is defined in Eq.~\eqref{eq:ConstG}.

\begin{figure}
\subfloat[]{\includegraphics[width=0.5\figurewidth]{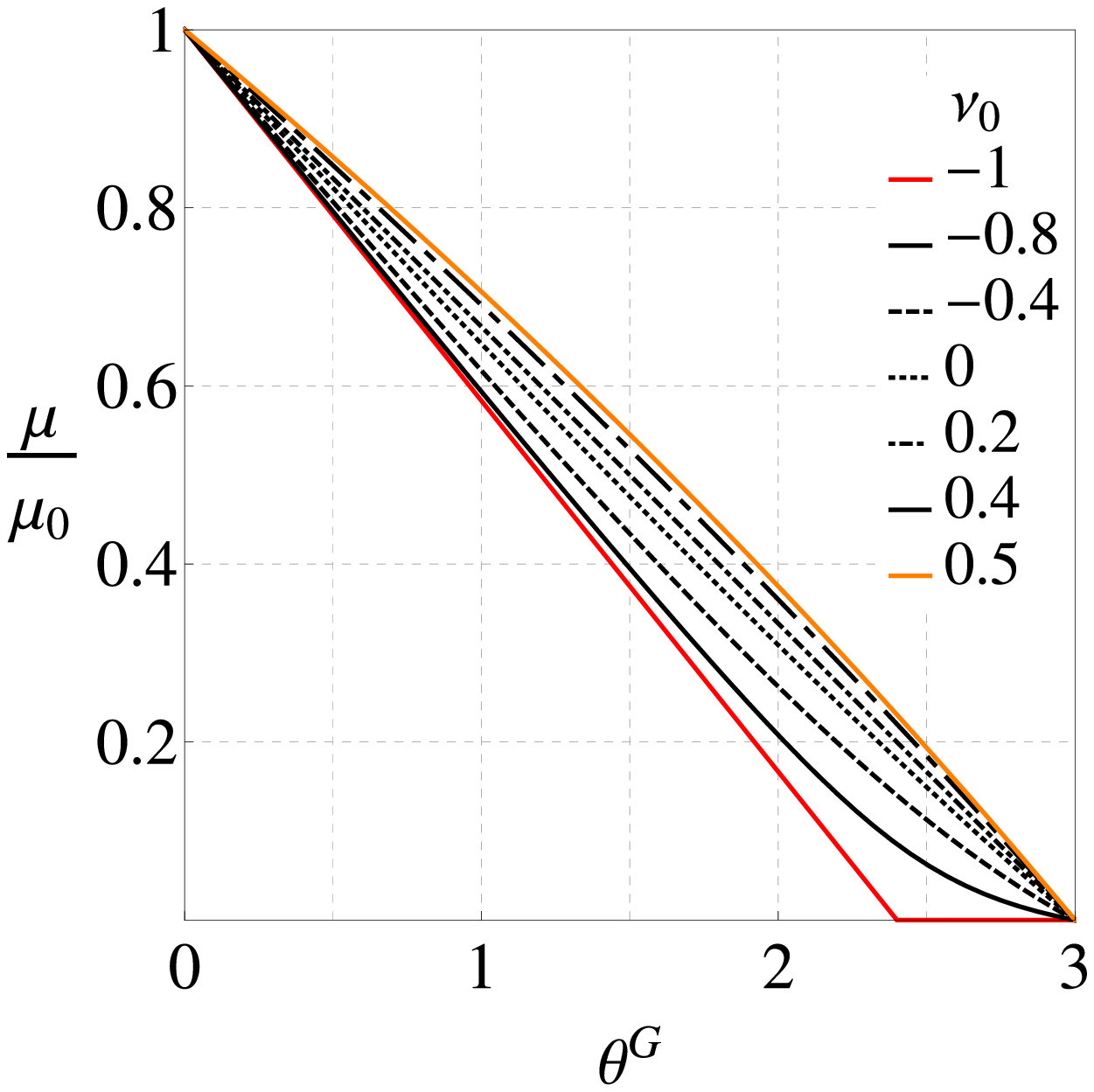}}
\subfloat[]{\includegraphics[width=0.5\figurewidth]{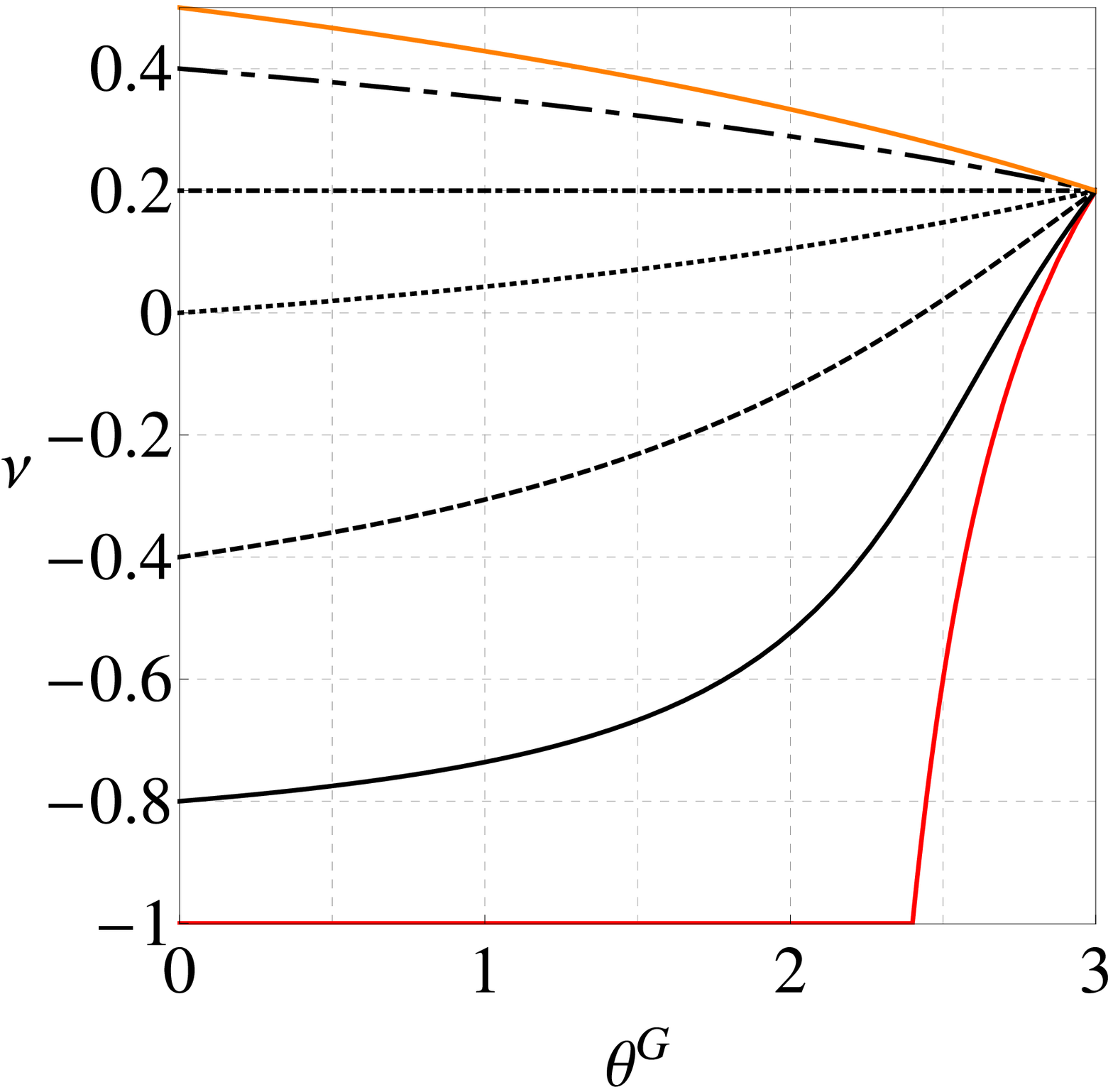}}
  \caption{
    \label{fig:MNforGconst} Constraint \eqref{eq:ConstG}: {\bf (a)}
    Renormalization $\mM/\mI$ of the shear modulus as a function of
    the dimensionless energy $\thetaG=2\beta\pEG$ of the built-in
    stress, for the constraint in Eq.~\eqref{eq:ConstG}. {\bf (b)} The
    effective Poisson ratio $\nM$ as a function of $\thetaG$ for
    several values of of the bare Poisson ratio $\nI$. In both panels,
    the orange lines correspond to the uniform-liquid limit for the
    bare medium, $\nI\to1/2$, while the red lines correspond to an
    infinitely compressible solid, $\nI\to-1$. The legends are
    identical in the two panels. }
\end{figure}

\begin{figure}
\includegraphics[width=1\figurewidth]{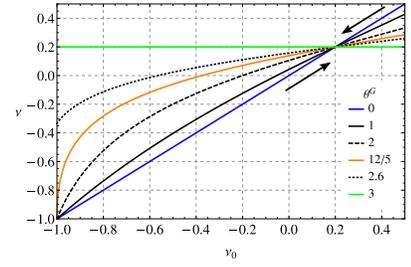}
\caption{
  \label{fig:NNforGconst} Constraint \eqref{eq:ConstG}:
  Dependence of the effective Poisson ratio $\nM$ on its bare value
  $\nI$ for several values of the built-in stress $\thetaG$. The three
  fixed points at $\nu = -1, 1/5, 1/2$ are discussed in text.  The
  arrows indicate the direction of the flow on $\nI\mapsto\nM$
  mapping.}
\end{figure}

\begin{figure} \vspace{10mm}
  \subfloat[]{\includegraphics[width=0.8\figurewidth]{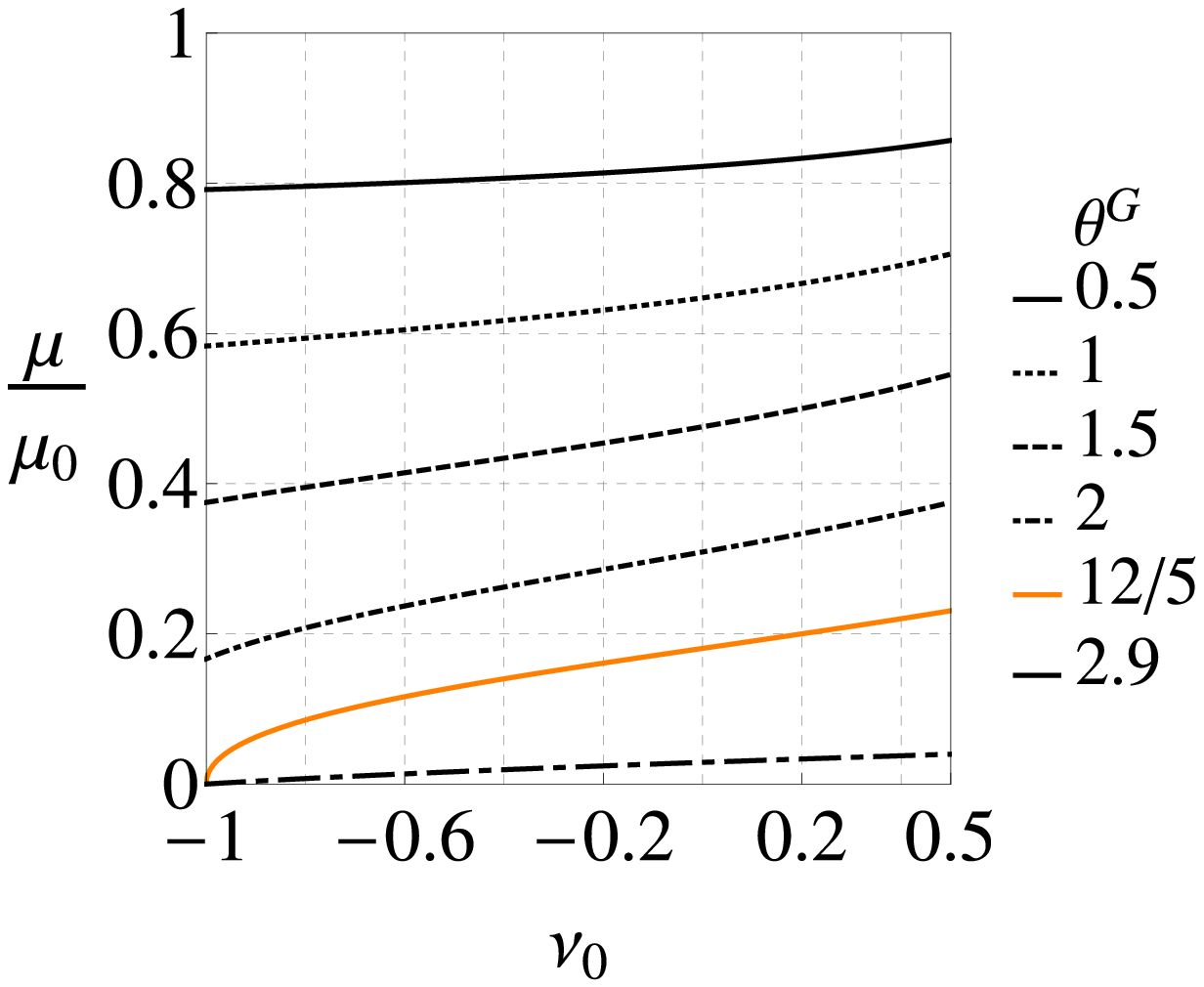}} 
 \\ 
  \subfloat[]{\includegraphics[width=0.8\figurewidth]{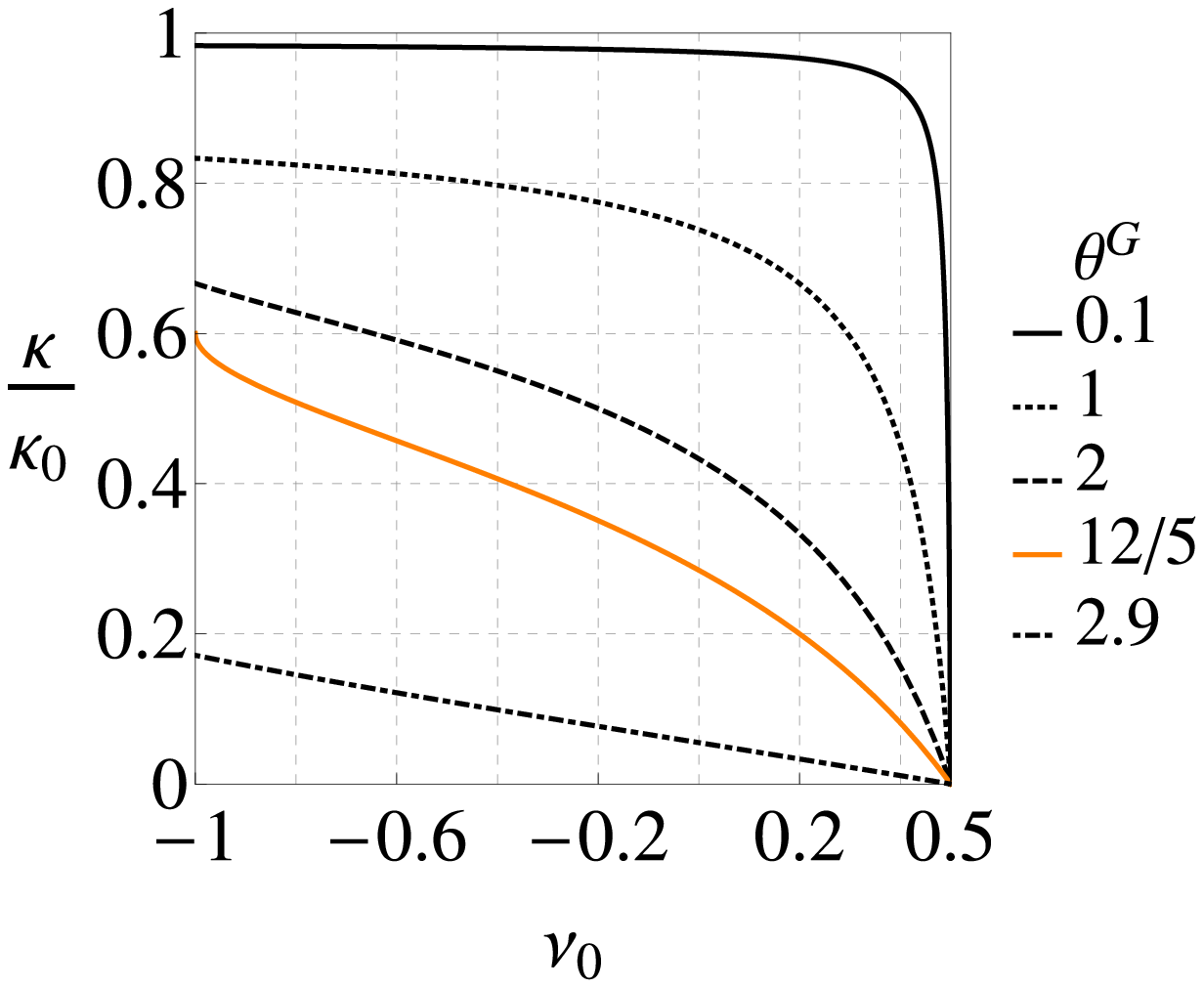}}
\caption{
  \label{fig:MKforGconst} Constraint \eqref{eq:ConstG}: Dependences of
  the ratios $\mM/\mI$ and $\kM/\kI$ on the bare value of the Poisson
  ratio $\nI$ for several values of $\thetaG$, in panels {\bf (a)} and
  {\bf (b)} respectively.  The orange line corresponds to
  $\thetaG=12/5$, which separates the two regimes in which the fixed
  point $\nu = -1$ is continuous and discontinuous respectively. Note
  that the derivative of $\mM/\mI$ as a function of $\nI$ diverges at
  $\thetaG=12/5$, $\nI\to-1$, whereby $\mM/\mI \propto \sqrt{(1 +
    \nI)}/2\sqrt{5}$. The $\kM/\kI$ ratio is zero at $\nu = 1/2$ for a
  finite $\thetaG$, however small. This ratio is finite when $\thetaG$
  is strictly zero.}
\end{figure}

The system of equations \eqref{eq:constG:Aeqs} can be readily solved,
the solution graphically summarized on
Figs.~\ref{fig:MNforGconst}-\ref{fig:MKforGconst}. This is the main
result of the present work, besides the formal developments in
Sections~\ref{sec:symm-betw-electr}-\ref{sec:cavity:gen} that lay
foundation of continuum mechanics for structurally degenerate solids.

It is immediately clear from Eq.~\eqref{eq:constG:Aeqs} that the
effective shear modulus is always reduced from its bare value in the
presence of built-in stress, since $-1 \le \nI, \nM \le 1/2$. This
down-renormalization comes about because the built-in stress enhances
the local elastic field, according to Eq.~\eqref{dF} and in contrast
with the dielectric case.

Because of the physical constraint $\mM\ge0$---which guarantees
mechanical stability with respect to shear, by
Eq.~\eqref{Felast}---the dimensionless built-in stress $\thetaG$ has a
limiting value: $\thetaG \le 3$. Beyond this limiting value of
built-in stress, the aperiodic solid becomes a uniform liquid.  The
dependences of the $\mM/\mI$ ratio and the Poisson ratio on
$\theta^\tsc{g}$ are shown in Fig.~\ref{fig:MNforGconst}(a) and (b)
respectively. When the compressibility diverges, $\nI\to-1$, the
$\mM/\mI$ ratio approaches the line $1-5\thetaG/12$, while in the
limit of uniform liquid $\nI\to1/2$, the ratio tends to the line
$6\lp3-\thetaG\rp/\lp18-\thetaG\rp$.

We have already discussed that the $\nI\mapsto\nM$ mapping has two
trivial fixed points corresponding to the uniform liquid ($\nM = \nI =
1/2$) and infinitely compressible solid ($\nM = \nI = -1$).  At the
uniform liquid fixed point, the bulk modulus vanishes at any value of
the built-in stress:
\begin{equation} \label{eq:constG:Klim2} \frac{\kM}{\kI} \,
  \xrightarrow[{\nI\to1/2}]{} \, 16 \left(\half - \nI \right)
  \frac{\lp6-\thetaG\rp\lp3-\thetaG\rp}{\thetaG\lp18-\thetaG\rp}.
\end{equation}
At the same time, the $\mM/\kM$ ratio remains finite in this limit,
except when $\thetaG \to 0$:
\begin{equation}
  \label{eq:constG:MKlim2}
  \lim_{\nI\to1/2}\frac{\mM}{\kM}=\frac{3\thetaG}{4\lp6-\thetaG\rp},
\end{equation}
or, equivalently, 
\begin{equation} \label{eq:nu}
  \lim_{\nI\to1/2} \nM =1+\frac{4}{\theta^\tsc{g} -8}.
\end{equation}
Because the renormalized $\mM/\kM$ ratio remains finite even as the
bare ratio vanishes---see Fig.~\ref{fig:NNforGconst}---the
uniform-liquid fixed point is {\em discontinuous}, except when there
is no built-in stress to begin with, $\thetaG = 0$.

The present formalism is internally-consistent in that it yields an
infinitely-compressible and, hence, marginally stable system, if one
supposes that a uniform liquid could sustain built-in stress of finite
magnitude, see Eq.~\eqref{eq:constG:Klim2}. In other words, we have
established that the internal stress is self-consistently zero in the
uniform-liquid regime. On the other hand, because only non-zero values
of the built-in stress are meaningful in the aperiodic-crystal state,
the discontinuity at $\nI = 1/2$ (for finite $\thetaG$) in
Eq.~\eqref{eq:nu} means that the RFOT transition from the uniform
liquid to the equilibrium aperiodic solid is discontinuous, while the
built-in stress also emerges at the transition in a discontinuous
fashion. In the RFOT theory, the discontinuity is signalled by a
discrete jump of the force constant for the effective Einstein
oscillator from zero to a number of order $100/a^2$.\cite{dens_F1,
  dens_F2, BausColot, Lowen, RL_LJ}

The $\nI=-1$ fixed point is continuous for sufficiently low values of
the built-in stress but becomes discontinuous when $\thetaG>12/5$,
where the discontinuity in the Poisson ratio is equal to:
\begin{equation}
  \label{eq:constG:Nlim1}
  \lim_{\nI\to-1}\nM=1-\frac{4}{5 (\theta -2)},
\end{equation}
In this regime, $\mM$ and $\kM$ vanish simultaneously while their
ratio remains finite, similarly to Eq.~\eqref{eq:constG:MKlim2}:
\begin{equation}
  \label{eq:constG:MKlim1}
  \lim_{\nI\to-1}\frac{\kM}{\mM}= -\frac{4 (5 \thetaG -12)} 
  {3 (5 \thetaG -18)},\QQText{for }\thetaG>\frac{12}{5}.
\end{equation}
The physical meaning of the $\nM = -1$ fixed is not entirely clear at
present. It may correspond to the mechanical stability limit of a
non-degenerate vibrational ground state. We anticipate that such a
stability limit could be realized in nature during pressure-induced
amorphization,\cite{Sharma1996, Brazhkin2003, Ponyatovsky1992,
  Hemley1988} which is possible when the crystalline structure is
relatively open. The latter situation may be also realized in high
spatial dimensions, where the highest possible filling fraction may be
achieved in {\em aperiodic} packings.\cite{RevModPhys.82.789}

In addition, according to Figs.~\ref{fig:MNforGconst}(b) and
\ref{fig:NNforGconst}, there is a non-trivial fixed point at
$\nI=\nM=\nI^\tsc{fp}=1/5$, independent of $\thetaG$. This fixed point
formally stems from the property of the Eshelby tensor, by which
$\tz{S}\propto\tz{I}$ for $\nM=1/5$, according to
Eq.~\eqref{eq:EshTens:fifth}.  Relation \eqref{eq:constG:Teq} then
dictates that $\cM\propto\cI$, which is possible only if
$\nI=\nM$. The proportionality of the Eshelby tensor to the unit
matrix means that at this special value of the Poisson ratio, the
relative weight of shear and uniform deformation of a relaxed
standalone inclusion does not change after it is inserted in the
matrix.  At this fixed point, the self-consistency relation
\eqref{eq:constG:Teq} boils down to a simple:
\begin{equation}
  \label{eq:constG:Teq:fifth}
  \mM^\tsc{fp}=\mI\lp1-\frac{\thetaG}{3}\rp.
\end{equation}
The non-trivial fixed point is {\em attractive}, because $\nM > \nI$
for $\nI < \nI^\tsc{fp}$, and $\nM < \nI$ for $\nI > \nI^\tsc{fp}$.
Note also that $\nM$ approaches $\nI^\tsc{fp}$ as $\thetaG$ tends to
its limiting value $3$, for all values of $\nI$.  Conversely, the {\em
  trivial} fixed points are unstable, as indicated by the arrows in
Fig.~\ref{fig:NNforGconst}.

The behavior of the effective moduli, in relation to their bare
counterparts, is shown in Fig.~\ref{fig:MKforGconst}. Here we
explicitly see that like the shear modulus, the bulk modulus is also
always down-renormalized. Finally, the renormalization flow in the
$(\mM, \kM)$ plane is shown in Fig.~\ref{fig:Renormalization} in
Appendix~\ref{app:flows}.

\subsection{Constraint 2: Built-in stress inserted in bare medium}
\label{sec:constF}

When written out explicitly, the constraint in
Eq.~\eqref{eq:ConstGlobal} reads as follows:
\begin{equation}\label{eq:Ep0}
  \frac{\pEF}{v}=\frac{1}{3\kI+4\mI}\lp\frac{2\mI}{9\kI}\fsC_{ii}^2
  +\frac{9\kI+8\mI}{20\mI}\,{}^\prime\!\fsC_{ij}^2\rp.
\end{equation} 
Note the {\em adiabatic} values of the elastic moduli satisfy the
relation $3\kM_s+4\mM_s=3\rho c_l^2$, where $\rho$ is the mass density
of the body and $c_l$ is the speed of longitudinal
phonons.\cite{LLelast} The expressions we have written for the elastic
free energy density are equilibrium, implying the elastic moduli are
isothermal. The isothermal and adiabatic shear moduli are equal to
each other\cite{LLelast}, while the adiabatic bulk modulus exceeds its
isothermal value, although usually not by much.

\begin{figure}
\subfloat[]{\includegraphics[width=\figurewidth]{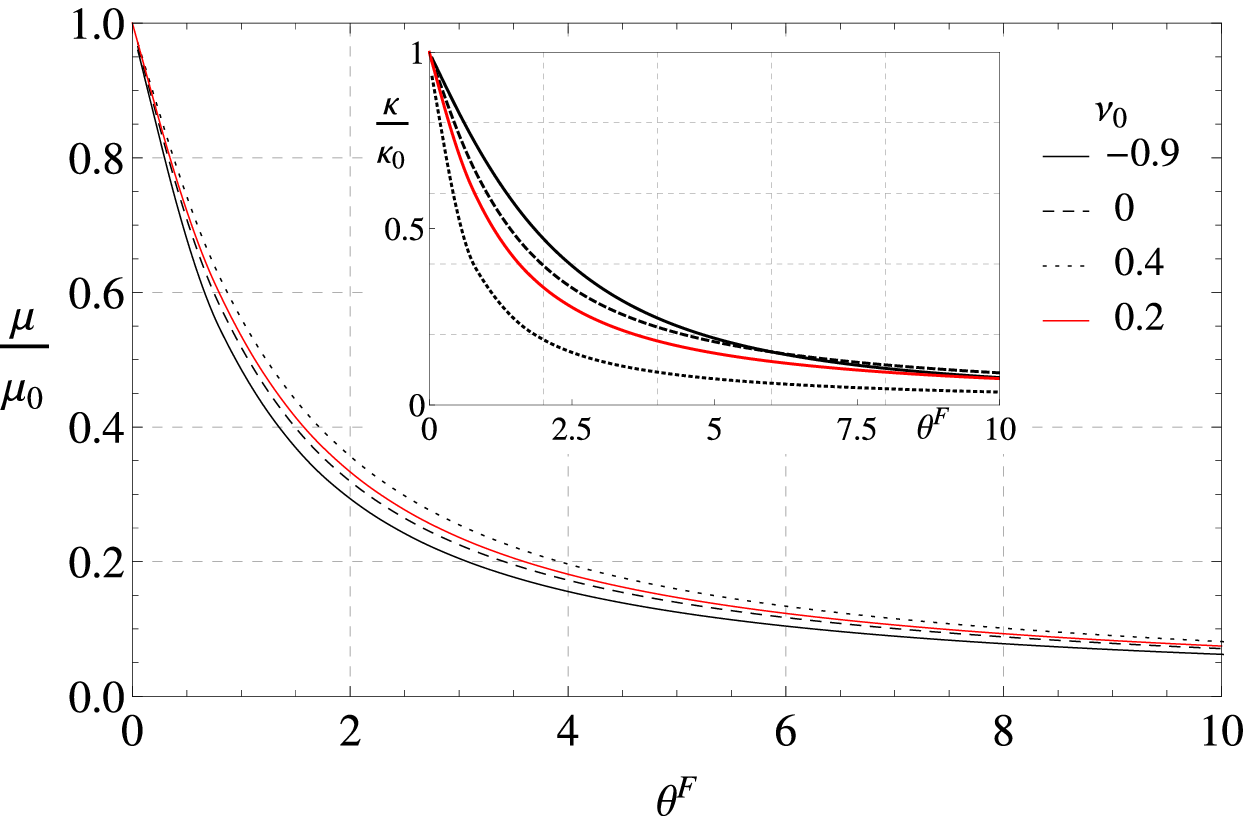}
}\\
\subfloat[]{\includegraphics[width=0.95\figurewidth]{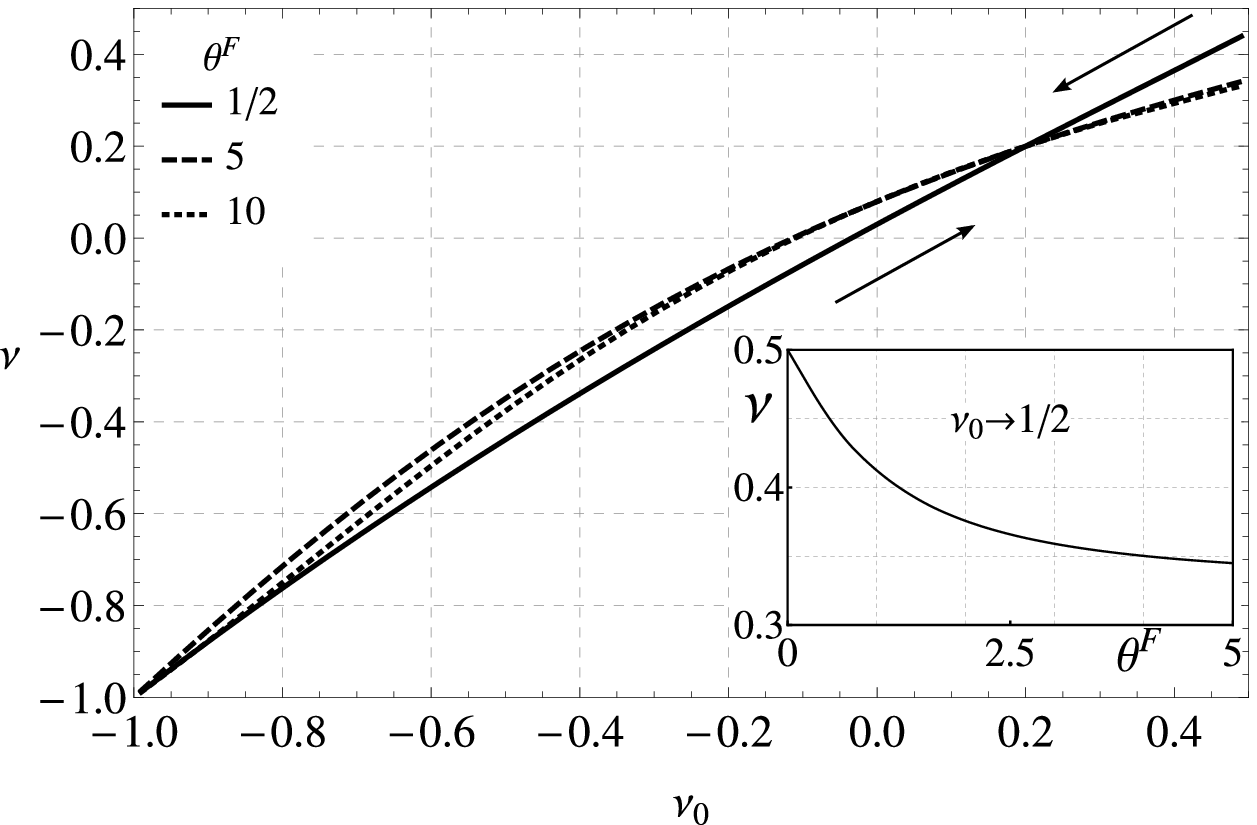}}
\caption{
  \label{fig:SolInclusion} Constraint \eqref{eq:ConstGlobal}:
  graphical summary of the solution of Eq.~\eqref{eq:cMcI}. {\bf (a)}
  Renormalization $\mM/\mI$ of the shear modulus as a function of the
  dimensionless stress energy $\thetaF=2\beta\pEF$ from
  Eq.~\eqref{eq:ConstGlobal} for several values of the bare Poisson
  ratio $\nI$. The black lines are numerical solutions of
  Eq.~\eqref{eq:cMcI}. The red line gives the analytical solution at
  the fixed point $\nM=\nI=1/5$, Eq.~\eqref{eq:mu:FixedPoint}. {\bf
    (b)} Dependence of the effective Poisson ratio $\nM$ on its bare
  value $\nI$ for several values of $\thetaF$.  The arrows indicate
  the direction of the flow on the $\nI\mapsto\nM$ mapping. The inset
  shows the dependence of $\nM$ on $\thetaF$ for $\nI\to1/2$.  }
\end{figure}

Similarly to the preceding case, we observe that the identity of the
built-in stress interpolates between the frozen-in shear and uniform
deformation as the Poisson ratio is varied from $-1$ to $1/2$.  The
relation between the bare and effective elastic moduli now contains
modified Bessel functions and is significantly more complicated than
in the preceding case, see Eq.~\eqref{eq:cMcI} in
Appendix~\ref{app:Z:response}. We present the numerical solution of
this equation in Fig.~\ref{fig:SolInclusion}.

According to Fig.~\ref{fig:SolInclusion}(a), the dependence of
$\mM/\mI$ on $\thetaF$ is affected by the value of $\nI$ only
weakly. The magnitude of the renormalization itself remains
significant, however there is no longer a limiting value to the
built-in stress. This seems consonant with the lower degree of
self-consistency in the current set-up, whereby the outside of the
cavity is no longer represented by the effective medium.

The $\thetaF = \text{Const}$ case exhibits the same three fixed points
as the preceding constraint, including their assignment in terms of
being attractive or repulsive, Fig.~\ref{fig:SolInclusion}(b).  In
contrast, only the uniform-liquid point is now discontinuous, the size
of the discontinuity vanishing in the absence of built-in stress. The
dependence of the jump of the Poisson ratio on the magnitude of
built-in stress is presented in the inset of
Fig.~\ref{fig:SolInclusion}(b).  If we invoke the notion from the
Introduction Section that the built-in stress must be finite in
magnitude, we again arrive at a result that the uniform liquid turns
into an aperiodic solid in a discontinuous fashion. Still, this result
is not as strong as in the Onsager limit, in which the finite jump in
the built-in stress itself, at the transition, was established
self-consistently.  Finally, it can be shown analytically that the
non-trivial fixed point is located, again, at $\nI^\tsc{fp} = 1/5$,
see Appendix~\ref{app:Z:response}.

\subsection{Constraint 3: BL model}
\label{sec:constS}

\begin{figure} 
\subfloat[]{\includegraphics[width=\figurewidth]{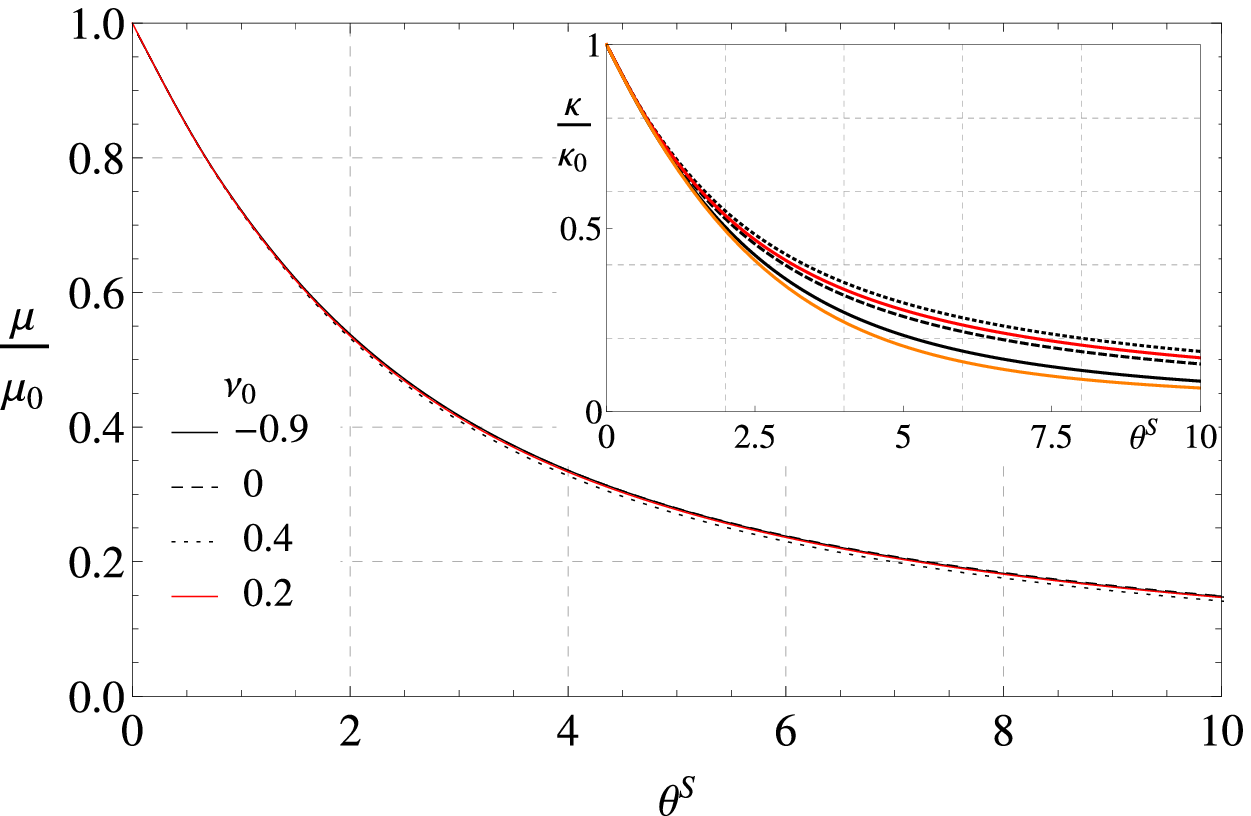}}\\
\subfloat[]{\includegraphics[width=0.95\figurewidth]{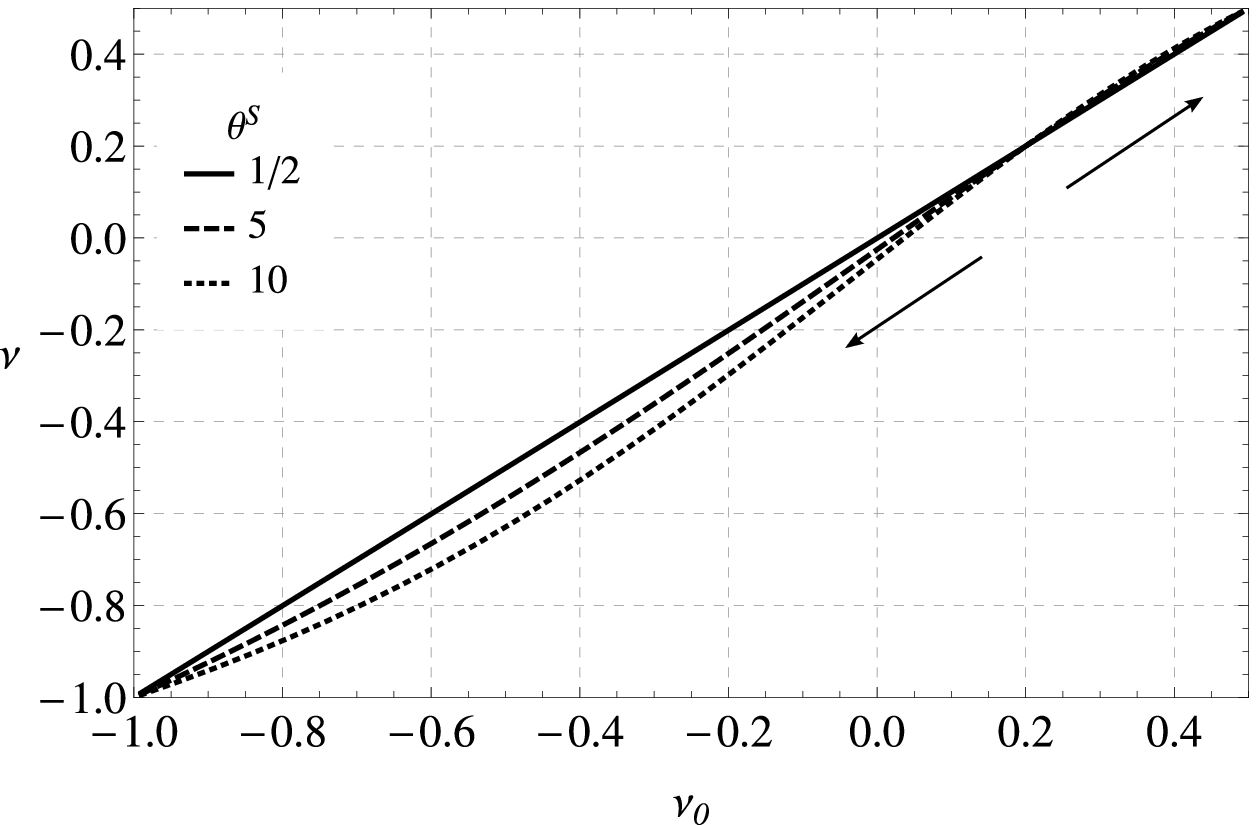}}
\caption{
  \label{fig:SolInclusion2} Constraint \eqref{eq:constraint:local}:
  graphical summary of the solution of Eq.~\eqref{eq:cMcI:Loc}.  {\bf
    (a)} Renormalization $\mM/\mI$ of the shear modulus as a function
  of the dimensionless stress energy $\thetaS=2\beta\pES$ from
  Eq.~\eqref{eq:constraint:local} for several values of the bare
  Poisson ratio $\nI$. The black lines are numerical solutions of
  Eq.~\eqref{eq:cMcI}. The red line gives the analytical solution at
  the fixed point $\nM=\nI=1/5$.  The orange line in the inset gives
  the analytical solution in the $\mI\to0$ limit,
  Eq.~\eqref{eq:Loc:nI12}.  {\bf (b)} Dependence of the effective
  Poisson ratio $\nM$ on its bare value $\nI$ for several values of
  $\thetaS$.  The arrows indicate the direction of the flow on the
  $\nI\mapsto\nM$ mapping. In contrast with the preceding cases, this
  fixed point is now unstable.  }
\end{figure}

The relation between the bare and renormalized moduli, which is given
as Eq.~\eqref{eq:cMcI:Loc:orig} in Appendix~\ref{app:Z:response}, also
must be solved numerically, see Fig.~\ref{fig:SolInclusion2}.

In contrast with the two preceding cases, the trivial fixed points are
now continuous in the full parameter range. Interestingly, the
relevance of the fixed points---in the RG sense of the word---is now
reversed.  The fixed point at $\nI=1/5$ is now repulsive, while the
trivial fixed points at $\nI=-1$ and $\nI=1/2$ are stable, see
Fig.~\ref{fig:SolInclusion2}(b). 

The repulsive nature of the $\nI^\tsc{fp} = 1/5$ point is consonant
with the BL finding that at the value of the Poisson ratio $1/5$, the
mean-field limit of the Hamiltonian (\ref{eq:Anzatz:Ham}) has a
continuous transition from an Ising-like ferromagnet to a
Heisenberg-like ferromagnet with 5-spins. The two regimes, when
well-developed, correspond to frozen-in uniform and frozen-in shear
stress patterns respectively. Exactly at the transition, the two types
of frozen-in stress patterns are marginally stable with respect to
each other. We thus conclude that the $\nI^\tsc{fp} = 1/5$ fixed point
is analogous to an isospin-like degeneracy between shear and uniform
deformation.

Note the uniform-liquid now transitions to the aperiodic solid {\em
  continuously}. This unphysical feature stems from the limitation of
the BL ansatz \eqref{eq:constraint:local} discussed earlier.

\section{Conclusion}
\label{sec:discussion}

We have developed a continuum mechanics description of the elasticity
exhibited by equilibrium, degenerate aperiodic solids. Such aperiodic
solids are exemplified by liquids that flow by local activated
transitions between distinct aperiodic free energy minima. The
transition to activated transport from the ordinary, collisional
transport typical of uniform liquids may occur above or below the
fusion temperature, depending on the fragility of the
liquid.\cite{LW_soft} In ordinary liquids---as opposed to colloids,
for instance---the structural glass transition is always preceded by
the emergence of activated transport.\cite{LW_Wiley}

Despite allowing the liquid to flow, albeit on very long timescales,
the activated transitions are rare events relative to molecular
vibrations.~\cite{LW_soft, LW_Wiley} Yet already a mesoscopic region
relaxes at a rate high enough to prevent one from defining a
vibrational ground state unambiguously in this region. At the same
time, the vibrational response of such a degenerate solid is well
described phenomenologically using the standard elasticity theory,
apart from the presence of a large dissipative component.

The present work shows why such an elastic description is possible
despite the lack of a unique vibrational ground state. Such uniqueness
is, of course, very basic to the theory of elasticity, similarly to
how the uniqueness of vacuum is basic to electrodynamics. To tackle
the problem of the vast structural degeneracy of equilibrium aperiodic
solids, we have employed the microscopic picture advanced by the RFOT
theory, which constructively demonstrates that such aperiodic crystals
represent a mosaic of distinct solutions of the free energy
functional.\cite{XW, KTW} The physical boundaries between distinct
solutions of the free energy functional are strained regions that
cannot be removed by elastic deformation, but only by a discontinuous
transition to the corresponding periodic crystal, if the latter
exists. The strained regions thus correspond to {\em built-in} stress.
The extent and concentration of such strained regions is dictated by
thermodynamics; near the glass transition, the corresponding
lengthscale is 2-4 nm in actual materials.\cite{XW, LW}

The specific implementation of the built-in stress employed in this
work originates from our earlier approach,\cite{BL_6Spin} in which the
molecular motions at short and long wavelengths are treated on a
separate footing. The short-wavelength modes give rise to a frozen-in
stress pattern, while the long-wavelength modes corresond to elastic
excitations of the material for a given configuration of the frozen-in
short-lengthscale motions.  Despite its tensorial nature, the
interaction between such local sources of strain bears similarities to
the electric dipole-dipole interaction. This deep analogy between
continuum electrostatics and mechanics allows one to formulate the
problem of interaction between sources of built-in stress similarly to
how Onsager~\cite{Onsager1936} and others\cite{debye12, debye29,
  kirkwood39} have derived the theory of dielectric response starting
from a microscopic model of a polar liquid as an equilibrium assembly
of molecular dipoles. In complete analogy with the dielectrics, the
mechanical response is determined by the elastic properties of
individual free energy minima (which corresponds to the response of
vacuum in electrodynamics), the magnitude of the built-in stress
(which corresponds to the molecular dipole moment), and temperature.
In the language of Yoshino and Mezard,\cite{PhysRevLett.105.015504,
  7644758720120607} the elastic properties of individual minima and
the macroscopic liquid correspond to inter- and intra-basin elastic
moduli.  In an important distinction, here we consider response at
{\em finite}, even if low, frequencies.

Yet in contrast with the dielectric case---whereby polarized dipoles
screen the external field---the elastic response is {\em enhanced} by
the sources of built-in stress. As a result, there is a liming value
to the built-in stress that can be supported by aperiodic crystal. In
actual substances, the magnitude of the built-in stress is determined
by the molecular interaction. Here, we have treated the magnitude of
built-in stress as a flexible parameter. We have found three fixed
points in the mapping between the bare and renormalized elastic
constants, the renormalization stemming from the presence and
relaxation of the built-in stress. Two of the fixed points correspond
to the uniform liquid, in which the shear modulus is identically zero,
and to an infinitely compressible solid. We find that the transition
from the uniform liquid to the equilibrium aperiodic crystal is
discontinuous, in agreement with earlier conclusions of the RFOT
theory.\cite{dens_F1, dens_F2, BausColot, Lowen, RL_LJ}

There is also a somewhat surprising fixed point at which the Poisson
ratio stays constant after the renormalization. The corresponding
value $\nM = 1/5$ is special in that it signals a degeneracy of sorts
between the shear and uniform deformation. If a spherical inclusion is
inserted in an elastic continuum, and both are characterized by this
particular value of the Poisson ratio, the relative weights of the
shear and uniform deformation inside and outside will be equal to each
other. Note that the equality of the Poisson ratios between two
different materials in contact implies that a single acoustic wave
will refract into a single wave.  In the absence of such equality, the
refraction will result in two waves because the transverse and
longitudinal sound will refract differently. The $\nM = 1/5$ fixed
point turns out to coincide with the critical point in the mean-field
limit of the BL model,\cite{BL_6Spin} in which the aperiodic solid
makes a transition between two types of frozen-in stress patterns
corresponding to shear and uniform deformation respectively.

As pointed out in the Introduction, the present work specifically
addresses the effects of structural {\em degeneracy} on the
vibrational response of an equilibrium aperiodic solid, as opposed to
effects of aperiodicity in a fully {\em stable} lattice. A stable
aperiodic lattice exhibits non-affine displacements and
spatially-heterogeneous elasticity, leading to Rayleigh scattering of
acoustic waves. These fascinating features of harmonic~(!) aperiodic
lattices have been proposed as the cause of the apparent excess of
vibrational states in glasses often referred to as the Boson Peak,
requiring however that the lattice be near its mechanical stability
limit.\cite{0295-5075-73-6-892, ParisiBP, 0295-5075-104-5-56001} In
the absence of such marginal stability, purely elastic scattering
seems too weak to account for the apparent magnitude of phonon
scattering at Boson Peak frequencies.~\cite{ACAnd_Phi, Joshi} In
contrast, the presence of structural degeneracy leads to an entirely
distinct, {\em resonant} type of phonon scattering.  The resonances
are due to local transitions between distinct free energy minima of
the aperiodic solid.~\cite{LW, LW_RMP} A RFOT-based analysis shows
these elastic resonances do account quantitatively for the apparent
magnitude of the heat capacity and phonon scattering both in the
temperature range corresponding to the Boson Peak~\cite{LW_BP, LW_RMP}
and, at the same time, at lower, sub-Kelvin temperatures where the so
called two-level systems~\cite{LowTProp} dominate the thermal
properties of the glass.~\cite{LW, LW_RMP}

It is hoped that despite some computational complexity, the present
description will help advance applications of the RFOT theory to
actual materials. The present description enables one to realize the
BL program of modeling activated transport in liquids via a spin model
on a fixed lattice, while not requiring the full knowledge of the
complicated, many-body interactions between actual molecules. Instead,
only the elastic constants and the bead size are needed as the
microscopic input. In this regard, it would be interesting to
investigate a case in which the bare elasticity is not isotropic.

\begin{acknowledgments}

  The authors thank Peter G. Wolynes for insight and valuable
  discussions.  This work has been supported by the National Science
  Foundation (CHE-0956127), the Alfred P. Sloan Research Fellowship,
  and the Welch Foundation Grant E-1765.

\end{acknowledgments}

\appendix 
\section{The Green tensor for a point stress source}
\label{sec:stress:source}
Consider an infinite medium with a distribution of body force
$\mathbf{f}\ofR$, which is non-zero over a finite domain. The
distribution produces an elastic response described by the deformation
field $\mathbf{u}$ satisfying the following boundary value problem,
\begin{equation}\label{eq:bvp:pforce}
  \begin{dcases}
    \sC_{ij,j}+f_i=0,\\
    \mathbf{u}\ofR\xrightarrow[r\to\infty]{}0,
  \end{dcases}
\end{equation}
where the stress $\s$ is related to the deformation $\mathbf{u}$ by
the constitutive relation $\s=\cI\e$ and the elastic strain $\e$ is
defined in Eq.~\eqref{eq:def}.  One can solve
Eq.~\eqref{eq:bvp:pforce} by Fourier transforming
$\mathbf{u}$,\cite{Teodosiu1982} the result given by
\begin{equation}
  \label{eq:def:u}
  u_i\ofR=\int\dV^\prime\gAC_{ij}(\mathbf{r}-\mathbf{r}^\prime)
  f_i\lp\mathbf{r}^\prime\rp,
\end{equation}
where the second-rank Green tensor $\gA$ for isotropic elasticity is
provided in Eq.~\eqref{eq:def:kelvin}.

The force balance for the built-in stress is
$\sC_{ij,j}+\fsC_{ij,j}=0$. Substituting this equation, together with
Eq.~\eqref{eq:bvp:pforce}, into Eq.~\eqref{eq:def:u} and integrating
by parts yields:\cite{Qu2006}
\begin{equation}
  \label{eq:u:fs}
  u_i\ofR=-\half\int\dV^\prime\lb\gAC_{ij,l^\prime}+\gAC_{il,j^\prime}\rb
  \fsC_{jl}\lp\mathbf{r}^\prime\rp.
\end{equation}
Differentiating the above equation w.r.t. $x_j$ and symmetrizing,
according to the definition of $\e$ from Eq.~(\ref{eq:def}), we obtain
that the strain field resulting from built-in stress $\fs$ can be
calculated using a Green's function which is essentially the second
derivative of the kernel $\gamma^\tsc{a}_{ij}$:
\begin{equation} \label{eq:strain:G}
  \eC_{ij}=\int\dV^\prime\tzC{G}^\tsc{a}_{ijml}
  (\mathbf{r},\mathbf{r}^\prime)\fsC_{ml}\lp\mathbf{r}^\prime\rp, 
\end{equation}
where
\begin{equation}
  \label{eq:def:G2}
  \begin{split}
    \tzC{G}^\tsc{a}&_{ijml}(\mathbf{r},\mathbf{r}^\prime)=
    -\frac{1}{4}\lp\delta_{ip}\frac{\partial}{\partial x_j} +
    \delta_{jp}\frac{\partial}{\partial
      x_i}\rp\\&\times\lp\delta_{ql}\frac{\partial}{\partial
      x^\prime_m} +\delta_{qm}\frac{\partial}{\partial
      x^\prime_l}\rp\gAC_{pq}(\mathbf{r},\mathbf{r}^\prime),
  \end{split}
\end{equation}
Noting that $\frac{\partial}{\partial
  x^\prime_i}=-\frac{\partial}{\partial x_i}$, when acting on a
function of $\mathbf{r}-\mathbf{r}^\prime$, one obtains
Eq.~\eqref{eq:coupling:reg}.

\section{The physical meaning of the Eshelby tensor}
\label{app:eshelby}
The Eshelby tensor $\tz{S}$ comes about in continuum mechanics as the
solution to the following problem.\cite{Eshelby1957} Consider an
infinite isotropic body $\Bo$ characterized by an elastic moduli
tensor $\tz{C}$. Suppose a local region, call it $\I$, undergoes a
martensitic or some other structural transformation, thus resulting in
a stress distribution $\fs$ due to the mismatch between the
transformed region and the matrix. This stress is incompatible; it is
a source of a body force which causes a compensating elastic strain
$\eS$ to appear in the surrounding elastic medium.  As we just saw in
Appendix \ref{sec:stress:source}, $\eS$ can be calculated using
Eq.~\eqref{eq:strain:G}, where the integration is now over $\I$ only,
since $\fs$ is zero outside. Further, if $\fs$ is uniform, it can be
moved outside the integral, and so $\eS$ is now related to the volume
average of $\tz{G}^\tsc{a}$ over $\I$:
\begin{equation}
  \label{eq:app:eshelby}
  \eSC_{ij}\ofR=\fsC_{ml}
  \int_\I\dV^\prime\tzC{G}^\tsc{a}_{ijml}(\mathbf{r}-\mathbf{r}^\prime).
\end{equation}
J. D. Eshelby showed that for an elliptical $\I$, the volume average
of $\tz{G}^\tsc{a}$ in (\ref{eq:app:eshelby}) does not depend on
$\mathbf{r}$, if $\mathbf{r}\in\I$. Thus, $\eS$ is homogeneous inside
$\I$.\cite{Eshelby1957} The Eshelby solution is usually written in
terms of the ``eigenstrain'' $\e^\ast$ related to $\fs$ by
\begin{equation}
  \label{eq:eigenstrain}
  \e^\ast=-\cM^{-1}\fs.
\end{equation}
The tensor relating the eigenstrain $\e^\ast$ with the actual elastic
strain $\eS$ \emph{inside} $\I$ is called the Eshelby tensor $\tz{S}$:
\begin{equation}
  \label{eq:def:eshelby:pure}
  \eS=\tz{S}
  \,\e^\ast.
\end{equation}
Eq.~\eqref{eq:app:eshelby} and \eqref{eq:eigenstrain} imply
\begin{align*}
  \eS&=-\lb\int_\I\dV^\prime\tz{G}^\tsc{a}\!\lp\mathbf{r}
-\mathbf{r}^\prime\rp\rb\cM\,\e^\ast,
\end{align*}
which by Eq.~\eqref{eq:def:eshelby:pure} yields
Eq.~\eqref{eq:def:EshelbyTz}.

\section{Calculation of the partition function and local
  susceptibility for uniformly distributed built-in stress}
\label{app:Z:response}

It is possible to formulate the linear elasticity so that rank-2
tensors are presented as six component vectors, $\eC_{ij} \rightarrow
\eC_\alpha$, $\alpha = 1, 2, 3, 4, 5, 6$. Specifically: $\eC_1 =
\eC_{11}$, $\eC_2 = \eC_{22}$, $\eC_3 = \eC_{33}$, $\eC_4 = \sqrt{2}
\eC_{23}$, $\eC_5 = \sqrt{2} \eC_{31}$, $\eC_5 = \sqrt{2}
\eC_{12}$.\cite{Voigt} To avoid confusion, the components of the
6-vectors will be labeled with Greek indexes. The original rank-4
tensors now become rank-2 tensors, as in
$\cM_{ijkl}\rightarrow\cM_{\alpha\gamma}$. Any isotropic tensor can be
diagonalized according to:
\begin{equation}\label{eq:def:decomp}
\Big(\Lt{a}{b}\Big)_{\alpha\beta}=\tzC{U}_{\alpha\gamma} 
\Big(\Dt{a}{b}\Big)_{\gamma\delta}\lp\tzC{U}^\top\rp_{\delta\beta},
\end{equation}
where
\begin{equation}\label{eq:def:U}
  \tz{U}=
  \left(
    \begin{array}{cccccc}
      \sfrac{1}{\sqrt{3}} & \sfrac{-1}{\sqrt{6}} & \sfrac{-1}{\sqrt{2}} & 0 & 0 & 0 \\
      \sfrac{1}{\sqrt{3}} & \sfrac{-1}{\sqrt{6}} & \sfrac{1}{\sqrt{2}} & 0 & 0 & 0 \\
      \sfrac{1}{\sqrt{3}} & \sfrac{\sqrt{2}}{\sqrt{3}} & 0 & 0 & 0 & 0 \\
      0 & 0 & 0 & 1 & 0 & 0 \\
      0 & 0 & 0 & 0 & 1 & 0 \\
      0 & 0 & 0 & 0 & 0 & 1
    \end{array}
  \right)
\end{equation}
is the tensor constructed from the eigenvectors of $\Lt{a}{b}$, and
the symbol $\Dt{a}{b}$ labels a diagonal tensor of the form
\begin{equation}\label{eq:def:D}
  \Dt{a}{b}=
  \left(
    \begin{array}{cccccc}
      a & 0 & 0 & 0 & 0 & 0 \\
      0 & b & 0 & 0 & 0 & 0 \\
      0 & 0 & b & 0 & 0 & 0 \\
      0 & 0 & 0 & b & 0 & 0 \\
      0 & 0 & 0 & 0 & b & 0 \\
      0 & 0 & 0 & 0 & 0 & b
    \end{array}
  \right).
\end{equation}
Note that $\tz{U}$ does not depend on $a$ and $b$,
$\tz{U}\tz{U}^\top=\tz{I}$, and the determinant $|\tz{U}|=-1$.

\subsection{Constraint 1, Eq.~\eqref{eq:ConstG}}

We begin with the first constraint, Eq.~\eqref{eq:ConstG}. Define a
6-vector $\ut$ such that
\begin{equation}\label{eq:constG:fs}
  \fs \equiv \sqrt{\frac{2\pE^\tsc{g}}{v}}\tz{G}^{-\sfrac{1}{2}}\ut.
\end{equation}
Substituting Eq.~\eqref{eq:constG:fs} into constraint
\eqref{eq:ConstG} we get in the $6$-vector representation:
\begin{equation}
  \label{eq:eF:tau}
          \utC_{ij}\utC_{ij}=\utC_\alpha\utC_\alpha=|\ut|^2=1. 
\end{equation}
Thus, the constraint~\eqref{eq:ConstG} is equivalent to fixing the
length of the $6$-vector $\ut$.  All structural states of a
homogeneous $\fs$ in the spherical cavity $\I$ allowed by
\eqref{eq:ConstG} are now mapped onto all possible orientations of the
unit $6$-vector $\ut$ analogously to how the configurations of a polar
molecule in a dielectric solution are mapped onto all possible
orientations of its dipole moment. The thermodynamic average
$\langle\fs\rangle$ can be computed via the thermodynamic average of
$\langle\ut\rangle$, cf. Eq.~\eqref{eq:onsager:d},
\begin{equation}\label{eq:tau:av}
  \langle\ut\rangle=\frac{\int \dO_6\ut e^{-\beta
      \pE}}{\int \dO_6 e^{-\beta
      \pE}},
\end{equation}
where the integration is carried out over the solid angle in the six
dimensional spherical coordinate system,
\begin{equation}\label{eq:SolidAngle6D}
\begin{split}
\dO_6=&\sin^4x_1\sin^3x_2\sin^2x_3\sin x_4 \\
&\times\dx_1\dx_2\dx_3\dx_4\dx_5 ,
\end{split}
\end{equation} 
$x_\alpha\in\lb0,\pi\rb$, for $\alpha<5$, and
$x_5\in\lb0,2\pi\rb$. 

The potential energy $\pE$ can be written in terms of $\ut$ in a form
completely analogous to the dielectric case, Eq.~\eqref{eq:onsager:W}:
\begin{equation}\label{eq:constG:pE}
\pE=\pE^\tsc{g}+\bm{\upzeta}\ut+\frac{v}{2}\eT\cM\lb\cI-\cM\rb\tz{Q}\eT,
\end{equation}
where
\begin{equation}
  \label{eq:constG:eT}
  \bm{\upzeta} \equiv \sqrt{2v\pE^\tsc{g}}\tz{C}\tz{Q}\tz{G}^{-\sfrac{1}{2}}\eT.
\end{equation}
Clearly, $\pE$ depends on the cosine of the angle between
$\bm{\upzeta}$ and $\ut$ only. Thus, analogously to the dielectric
case, the partition function can be calculated exactly:
\begin{equation}\label{eq:constG:Z}
  \begin{split}
    Z&=\int\dO_6 e^{-\beta\pE}= \\&= \frac{8\pi^2}{3} e^{-\beta
      \pE^\tsc{g}}
    e^{-\frac{v}{2}\beta\eT\cM\lb\cI-\cM\rb\tz{Q}\eT}\int_0^\pi \dx_1
    \sin^4x_1e^{-y\cos x_1} \\&= 8\pi^3\frac{I_2(y)}{y^2} e^{-\beta
      \pE^\tsc{g}} e^{-\frac{v}{2}\beta\eT\cM\lb\cI-\cM\rb\tz{Q}\eT},
  \end{split}
\end{equation}
where $y\equiv\beta|\bm{\upzeta}|$ and we have used the integral
representation of the modified Bessel function $I_n(y)$ from
Eq.~(9.6.18) of Ref.~\onlinecite{AS}. Such integrals often appear in
the context of the $O(n)$ model.~\cite{Mussardo2010}

To compute the expectation value of the internal stress we first note
that by Eq.~\eqref{eq:EpConst}:
\begin{equation}\label{eq:constG:F1}
  \begin{split}
    \frac{\partial F}{\partial\eTC_\alpha}& =
    -\frac{1}{\beta}\frac{1}{Z}\frac{\partial Z}{\partial\eTC_\alpha}\\
    &=\lp v \tz{C}\tz{Q}\la\fs\ra\rp_\alpha+\lp v
    \cM\lb\cI-\cM\rb\tz{Q}\eT\rp_\alpha.
  \end{split}
\end{equation}
One the other hand, differentiation of Eq.~\eqref{eq:constG:Z} yields
\begin{equation}\label{eq:constG:F2}  \frac{\partial F}{\partial\eTC_\alpha}=
  v\lp\cM\lb\cI-\cM\rb\tz{Q}\eT\rp_\alpha - \frac{1}{\beta} 
  \frac{I_3(y)}{I_2(y)}\frac{\partial y}{\partial \eTC_\alpha}, 
\end{equation}
where 
\begin{equation} \frac{\partial y}{\partial \eTC_\alpha}=2\beta^2 v
  \frac{\pE^\tsc{g}}{y}\lp\cM^2\tz{Q}^2\tz{G}^{-1}\eT\rp_\alpha.
\end{equation}
Here we have used Eq.~(9.6.28) of Ref.~\onlinecite{AS}. Combining
Eqs.~\eqref{eq:constG:F1} and \eqref{eq:constG:F2} yields
\begin{equation}\label{eq:constG:fs2}
  \begin{split}
    \la\fsC_\alpha\ra&=-\frac{1}{\beta v}\lp\cM\tz{Q}
    \rp^{-1}_{\alpha\gamma}\frac{I_3(y)}{I_2(y)}
    \frac{\partial y}{\partial \eTC_\gamma}    \\
    &=-2\beta\pE^\tsc{g}\frac{I_3(y)}{y
      I_2(y)}\lp\cM\tz{Q}\tz{G}^{-1}\eT\rp_\alpha.
  \end{split}
\end{equation}
Note that $\lim_{y\to0} I_3(y)/(y I_2(y))=\frac{1}{6}$.  Upon
differentiation of Eq.~\eqref{eq:constG:fs}, we obtain for the static
susceptibility
\begin{equation}
\begin{split}\label{eq:constG:sus}
  \susC^\tsc{\,s}_{\alpha\gamma}
=-\frac{\beta\pE^\tsc{g}}{3}\lp\cM\tz{Q}\tz{G}^{-1}\rp_{\alpha\gamma},
\end{split}
\end{equation}
Combining this with Eqs.~\eqref{eq:ConstG}, \eqref{eq:def:G}, and
\eqref{eq:def:CM3} yields Eq.~\eqref{eq:constG:Teq}.

\subsection{Constraint 2, Eq.~\eqref{eq:ConstGlobal}}

Now we define the
vector $\ut$ according to:
\begin{equation}\label{eq:TauDef}
\fs = \sqrt{\frac{2\pEF}{v}}\tz{R}\;\tz{U}\;\ut,
\end{equation}
where
\begin{equation}\label{eq:def:R}
\begin{split}
  \tz{R} & \equiv \sqrt{\cI\lp\tz{I}-\tz{S}_0\rp^{-1}}=\sqrt{\lp3\kI+4\mI\rp}\\
  &\times\Lt{\sqrt{\frac{3\kI}{4\mI}}}{\sqrt{\frac{10\mI}{9
        \kI+8\mI}}}
\end{split}
\end{equation}
and the square root of the tensor is computed using
Eq.~\rf{eq:L:power}.

With these definitions, the potential energy $\pE$ from
Eq.~\rf{eq:EpConst} can be written as:
\begin{align}\label{eq:EpTau}
\frac{\pE}{v} &= \frac{\pEF}{v} \ut\; \tz{D}\;\ut + \sqrt{\frac{2\pEF}{v}}
\eT\tz{C\;Q\;R}\;\tz{U}\ut\notag\\
  &- 
\frac{1}{2}\eT\cM\lb\cM-\cI\rb \tz{Q}\; \eT,
\end{align}
where
$\tz{D}=\tz{U}^\top\tz{RGRU} =\Dt{\gamma_1}{\gamma_2}$ 
is a diagonal matrix, see \rf{eq:def:D}, with 
\begin{align} 
  \gamma_1&=\frac{3 \smr  (1-\nI)}{1+2\smr(1-2\nI)+\nI},\notag\\
  \gamma_2&=\frac{15 \smr (1-\nI) (7-5\nM)}{\lp7-5\nI\rp \lp8+7
    \smr-5\nM\lb2+\smr\rb\rp}.
\end{align}

Eqs.~\eqref{eq:cavity:susceptibility} and \eqref{eq:TauDef} yield:
\begin{equation}
  \label{eq:static:susceptibility:6vector}
  \begin{split}
    \!\tzC{X}^\tsc{s}_{\alpha\gamma}
    =
\sqrt{\frac{2\pEF}{v}}\lp\tz{R}\tz{U}\rp_{\alpha\zeta}\frac{\partial\la\utC_\zeta\ra}{\partial\eTC_{\gamma}}\Bigg|_{\eT=0},
\end{split}
\end{equation}
where the derivative on the r.h.s. can be rewritten, using
Eqs.~\eqref{eq:tau:av} and \eqref{eq:EpTau}, as
\begin{align}\label{eq:av:tau2}
  \frac{\partial\langle\utC_\alpha\rangle}{\partial\eTC_\gamma}\Bigg|_{\eT=0}&=
  -\beta \sqrt{\frac{2\pEF}{v}} \Bigg[ \langle
  \utC_\alpha\utC_\zeta\rangle-\langle
  \utC_\alpha\rangle\langle\utC_\zeta\rangle
  \Bigg]_{\eT=0}\notag\\
&\times\lp \tz{U}^\top\tz{R}\;\tz{Q\;C}\rp_{\zeta\gamma}.
\end{align}
The angular integration in Eq.~\eqref{eq:tau:av} can be performed
analytically since for $\eT=0$,
\begin{align}\label{eq:E:et0}
&\frac{\pE}{\pEF}\Bigg\vert_{\eT=0}
  =\ut\tz{D}\ut\notag\\&=\frac{1}{2}\lb\gamma_1+\gamma_2+\lp\gamma_1-\gamma_2\rp \cos2x_1\rb,
\end{align}
where $x_1$ is from Eq.~\eqref{eq:SolidAngle6D}. By symmetry,
$\langle\utC_\alpha\rangle_{\eT=0}=0$.  A straightforward calculation
yields
\begin{equation}\label{eq:av:tau3}
  \langle
  \utC_\alpha\utC_\zeta\rangle\Bigg|_{\eT=0}
  =\Big(\Dt{1-5s(p)}{s(p)}\Big)_{\alpha\zeta},
\end{equation}
where
\begin{equation}\label{eq:def:pPar}
  p \equiv \frac{1}{4}\thetaF \lp\gamma_1-\gamma_2\rp.
\end{equation}
and
\begin{align} \label{eq:s(p)} s\lp p\rp&=\frac{\int_0^\pi dx_1
    \sin^6x_1e^{-p \cos2x_1}}{\int_0^\pi dx_1\sin^4x_1e^{- p
      \cos2x_1}}\\&=\frac{ p\lp4p-1\rp I_0(p)+\lp2+p\lb4p-3\rb\rp
    I_1(p)} {10p\lb 2p I_0(p)+\lb2p-1\rb I_1(p)\rb}. \nonumber
\end{align}

Substitution of these results into
Eq.~\eqref{eq:static:susceptibility:6vector} gives:
\begin{equation}\label{eq:def:CMresult1}
\tz{X}^\tsc{s}=-\thetaF\,\cI\cM\lp\tz{I}-\tz{S}_0\rp^{-1}\tz{Q}\Lt{1-5s(p)}{s(p)},
\end{equation} 
where $\thetaF=2\beta\pEF$, Eq.~\eqref{eq:ConstGlobal}.  Subsequently,
\begin{equation}\label{eq:def:CMresult2}
\begin{split}
\lp\tz{I}-\tz{S}_0\rp&
\lp\cM+\lb \cI-\cM\rb\tz{S}\rp
\lp\cI-\cM\rp\\
&=\thetaF\,\cI\cM\,\Lt{1-5s(p)}{s(p)}. 
\end{split}
\end{equation} 
Dividing out both sides of the last equation by $\cI^2$ yields the
sought relation between the bare and renormalized elastic moduli:
\begin{equation}\label{eq:cMcI}
\begin{split}
\lp\tz{I}-\tz{S}_0\rp&
\lp\MR+\lb\tz{I}-\MR\rb\tz{S}\rp
\lp\tz{I}-\MR\rp\\
&=\thetaF\;\MR\,\Lt{1-5s(p)}{s(p)},
\end{split}
\end{equation}
where 
\begin{equation}
  \label{eq:def:X}
  \MR  \equiv \cM\cI^{-1}=\smr\Lt{\frac{1+\nM}{1-2\nM}\frac{1-2\nI}{1+\nI}}{1}. 
\end{equation}
As discussed in Sec.~\ref{sec:symm-betw-electr}, the tensorial
Eq.~\eqref{eq:cMcI} is equivalent to a system of two scalar equations.
Analytical solution of Eq.~\eqref{eq:cMcI} is possible in the high
temperature limit, where $\Lt{1-5s(p)}{s(p)}=\tz{I}/6$. However, the
resulting expression is too bulky to give here.

The non-trivial fixed point can be found analytically usign the high
temperature limit. Taylor-expanding $s(p)$ from Eq.~\eqref{eq:s(p)}:
$s(p)=\frac{1}{6}+\frac{p}{72}+O\left(p^2\right)$, yields, together
with Eq.~\eqref{eq:cMcI}:
\begin{equation}
  \label{eq:cMcI:highT}
6\lp\tz{I}-\tz{S}_0\rp
\lp\MR+\lb\tz{I}-\MR\rb\tz{S}\rp
\lp\tz{I}-\MR\rp
=\thetaF\;\MR.
\end{equation}
To test for the presence of the fixed point we substitute
$\nM=\nI$. Then, $\MR=\smr\tz{I}$, $\tz{S}=\tz{S}_0$, and the equation
simplifies to read
\begin{eqnarray}
  \label{eq:cMcI:highT:nInM}
  &6\lp\tz{I}-\tz{S}_0\rp
  \lp\smr+\lb1-\smr\rb\tz{S}_0\rp
  \lp1-\smr\rp
  \nonumber \\ &=\thetaF\smr\tz{I}. 
\end{eqnarray}
Since the r.h.s. of Eq.~\eqref{eq:cMcI:highT:nInM} is proportional to
$\tz{I}$, the solution is possible only if $\tz{S}_0$ is proportional
to $\tz{I}$ as well, and so, by Eq.~\eqref{eq:S:sphere} with
$\nM=\nI$:
\begin{equation}\label{eq:OneFaith}
\nI=\nI^\tsc{fp}=\frac{1}{5},
\end{equation}
Remarkably, at the fixed point
$(\gamma_1-\gamma_2)_{\nM=\nI=\nI^\tsc{fp}}
=0\notag$,
so that 
\begin{equation}
  \label{eq:cMcI:rhs:FixedPoint}
  \Lt{1-5s(p)}{s(p)}\Big\vert_{\nM=\nI=\nI^\tsc{fp}}=\frac{1}{6}\tz{I},
\end{equation}
see Eq.~\eqref{eq:cMcI}. This means that at the fixed point,
Eqs.~\eqref{eq:cMcI:highT:nInM} and \eqref{eq:OneFaith}---which were
obtained by Taylor-expanding with respect to $p$ from
Eq.~\eqref{eq:def:pPar}---are valid for all values of $\thetaF$. Then,
selecting the positive root of Eq.~\eqref{eq:cMcI:highT:nInM} at the
fixed point, we obtain,
\begin{equation}
  \label{eq:mu:FixedPoint}
  \mM^\tsc{fp}=\lp\sqrt{1+\lb\frac{\thetaF}{3}\rb^2}-\frac{\thetaF}{3}\rp\;\mI.
\end{equation}
Since the the $\smr$ ratio depends on $\nI$ only weakly, see
Fig.~\ref{fig:SolInclusion}, Eq.~\eqref{eq:mu:FixedPoint} represents a
good approximate expression for the temperature dependence of the
effective shear modulus for materials with $\nI$ numerically close to
0.2.

\subsection{Constraint 3, Eq.~\eqref{eq:constraint:local}}

The preceding calculation is easily adopted for the constraint
\eqref{eq:constraint:local} by substituting for the matrix $\tz{R}$,
Eq.~\eqref{eq:def:R}, the following matrix:
\begin{equation}
  \label{eq:def:R:local}
\widetilde{\tz{R}}=\cI^{1/2},
\end{equation}
so we can switch to the constraint \eqref{eq:constraint:local} by
simply removing the factor $\tz{I}-\tz{S}_0$ from the formulas. This
yields the following relation between the bare and renormalized
moduli:
\begin{align} 
\lp\cM-\lb \cM-\cI\rb\tz{S}\rp&\label{eq:cMcI:Loc:orig}
\lp\cM-\cI\rp\\
&=-\thetaS\,\cI\cM\,\Lt{1-5s(\tilde{p})}{s(\tilde{p})}\notag
\end{align}
or
\begin{align}
  \lp\MR+\lb\tz{I}-\MR\rb\tz{S}\rp\label{eq:cMcI:Loc} \lp\tz{I}-\MR\rp
  &=\thetaS\;\MR\,\Lt{1-5s(\tilde{p})}{s(\tilde{p})},
\end{align}
where 
\begin{eqnarray}
  \label{eq:pPar:Loc}
\tilde{p}=&
(3/4)\smr\thetaS\lp3-5\nM+\nI\lb15\nM-13\rb\rp  \nonumber \\
\times& (1+\nI+2\smr\lb1-2\nI\rb)^{-1} \nonumber  \\
\times& (8+7\smr-5\nM\lb\smr+2\rb)^{-1}.
\end{eqnarray}

Similarly to Eq.~\eqref{eq:cMcI}, one can solve
Eq.~\eqref{eq:cMcI:Loc} in the high temperature limit, but the
resulting expression is very complicated; we only show it graphically
in Fig.~\ref{fig:SolInclusion2}. Note that unlike
Eq.~\eqref{eq:def:CMresult2}, Eq.~\eqref{eq:cMcI:Loc:orig} possesses
three {\em continuous} fixed points. Indeed, if $\nI=1/2$ then
$\cI\propto\tz{J}$, see Eq.~\eqref{eq:c:general}. Consequently, the
r.h.s.  of Eq.~\eqref{eq:cMcI:Loc:orig} is proportional to
$\tz{J}$. Hence, the l.h.s. must be proportional to $\tz{J}$ as well,
which is possible only if $\nM=1/2$. The solution of
Eq.~\eqref{eq:cMcI:Loc:orig} in this case is
\begin{equation}
  \label{eq:Loc:nI12}
\kM\Big\vert_{\mI=0}=\frac{\kI}{6}\left(\thetaS\lb  \frac{I_0\left(\thetaS/4\right)}{I_1\left(\thetaS/4\right)}-1\rb-2\right)
\end{equation} 
and is shown in the inset of
Fig.~\ref{fig:SolInclusion2}(a). Similarly, in the other extreme of
$\mI=-1$, $\cI\propto \tz{K}$, and so by Eq.~\eqref{eq:cMcI:Loc:orig},
$\cM$ must be proportional to $\tz{K}$,  leading to $\kM=0$.

In the high temperature limit, where Eq.~\eqref{eq:cMcI:Loc} reduces
to
\begin{equation}\label{eq:cMcI:Loc:highT}
6\lp\MR+\lb\tz{I}-\MR\rb\tz{S}\rp
\lp\tz{I}-\MR\rp
=\thetaS\MR,
\end{equation}
cf. Eq.~\eqref{eq:cMcI:highT}. If $\nM=\nI$, $\MR=\smr\tz{I}$, and
$\tz{S}=\tz{S}_0$. As a result, $\tz{S}_0$ must be proportional to
$\tz{I}$, leading to $\nI^\tsc{fp}=1/5$,
cf. Eq.~\eqref{eq:cMcI:highT:nInM}.  Also, $\tilde{p}=0$ at the fixed
point. Thus the analog of Eq.~\eqref{eq:mu:FixedPoint} for the third
constraint is given by
\begin{equation}
  \label{eq:mu:FixedPoint:Loc}
\mM^\tsc{fp}=\lp\sqrt{1+\lb\frac{\thetaS}{6}\rb^2}-\frac{\thetaS}{6}\rp\;\mI.
\end{equation}

\section{Renormalization flows in the elastic moduli space and
  positive-definiteness of the cavity energy function}
\label{app:flows}

Here we consider the renormalization flow on the
$(\mI,\kI)\mapsto(\mM,\kM)$ mapping for Eqs.~\eqref{eq:constG:Teq},
\eqref{eq:def:CMresult1} and \eqref{eq:cMcI:Loc:orig}. The mappings
depend on the dimensionless temperatures: $\thetaG$, $\thetaF$ and
$\thetaS$ respectively. Here we assume these three parameters are
small, with the aim of obtaining a continuous RG flow.  The linearized
mapping for Eq.~\eqref{eq:constG:Teq} corresponding to constraint
\eqref{eq:ConstG} can be obtained analytically. For the other two
equations, we rearrange them to the form $\Lt{f_1}{f_2}=0$, and then
Taylor expand near the solution to linearize the mapping and connect
small increments of the bare and effective elastic moduli, via
\begin{equation}
  \label{eq:Ren:Flow}\lp
  \begin{matrix}
    \delta\mM\\\delta\kM
  \end{matrix}
\rp=N(\cM,\cI,\theta)
\lp
  \begin{matrix}
    \delta\mI\\\delta\kI
  \end{matrix}
\rp,
\end{equation}
where the dimensionaless stress energy $\theta$ is set equal to its
values corresponding to the three constraints. The matrix $N$ has the
form
\begin{equation}
  \label{eq:def:RenMatrix}
N=\frac{1}{\Delta}\lp
\begin{array}{c|c}
  \frac{\partial f_1}{\partial \mI}\frac{\partial f_2}{\partial \kM}-\frac{\partial f_1}{\partial\kM} \frac{\partial f_2}{\partial\mI} & \frac{\partial f_1}{\partial \kI}\frac{\partial f_2}{\partial \kM}-\frac{\partial f_1}{\partial\kM} \frac{\partial f_2}{\partial\kI} \\[5pt]
  \hline~&~\\[-7pt]
  \frac{\partial f_1}{\partial \mM}\frac{\partial f_2}{\partial \mI}-\frac{\partial f_1}{\partial\mI} \frac{\partial f_2}{\partial\mM} & \frac{\partial f_1}{\partial \mM}\frac{\partial f_2}{\partial \kI}-\frac{\partial f_1}{\partial\kI} \frac{\partial f_2}{\partial\mM}  
\end{array}
\rp,
\end{equation}
\begin{equation} \label{eq:def:delta} \Delta=\frac{\partial
    f_1}{\partial \kM}\frac{\partial f_2}{\partial \mM}-\frac{\partial
    f_1}{\partial\mM} \frac{\partial f_2}{\partial\kM}.
\end{equation}

\begin{figure}
\subfloat[]{\includegraphics[width=0.56\figurewidth]{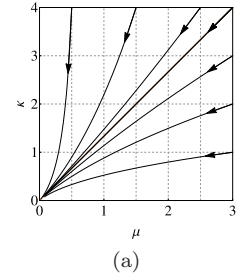}}\\
\subfloat[]{\includegraphics[width=0.6\figurewidth]{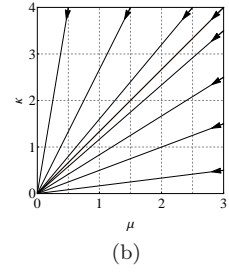}}
\caption{
  \label{fig:Renormalization} Renormalization flow for the
  $(\mI,\kI)\mapsto(\mM,\kM)$. The respective curvatures of the flow
  lines have opposite signs in panels (a) and (b).  Both flows are
  calculated for $\theta=0.1$.}
\end{figure}

The renormalization flows corresponding to constraints
Eq.~\eqref{eq:ConstG} and \eqref{eq:constraint:local} are shown in
Fig.~\ref{fig:Renormalization} (a) and (b) respectively. The flow for
the second constraint, Eq.~\eqref{eq:ConstGlobal}, looks very similar
to Fig.~\ref{fig:Renormalization}(a) and is not provided.  The
renormalization flow clearly reflects the down-renormalization of the
elastic constants, discussed in the main text, and leads toward the
state with $\mM=\kM=0$. This state formally corresponds to an
infinitely compressible, uniform liquid.

\begin{figure}
\includegraphics[width=0.5\figurewidth]{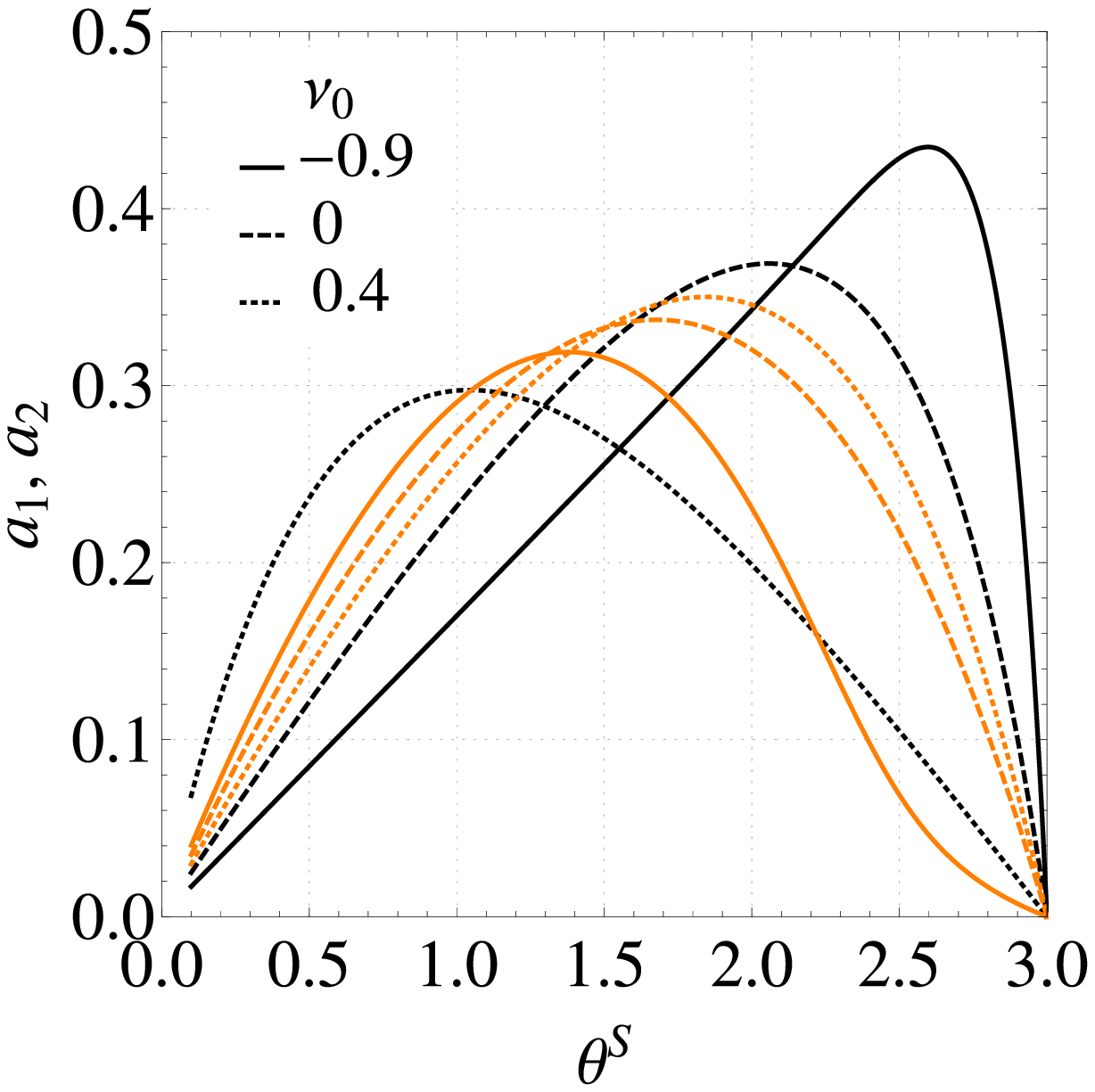}
\caption{
  \label{fig:TzA} Coefficients of the tensor $\tz{A}$,
  Eq.~\eqref{eq:def:TzA} corresponding to the solution of
  Eq.~\eqref{eq:constG:Teq}, for several values of the bare Poisson
  ratio. Black lines correspond to $a_1$, while the orange lines
  correspond to $a_2$, see text for explanation.}
\end{figure}

Finally, we show that the last term in the Eq.~\eqref{eq:EpConst},
\begin{equation}
  \label{eq:LastTerm}
  \frac{1}{2}\eT\cM\lp\cI-\cM\rp \tz{Q}\; \eT,
\end{equation}
which describes the cavity contribution to the potential energy of the
external load, is always positive.  First note that the tensor product
in~\eqref{eq:LastTerm} can be written as
\begin{equation}
  \label{eq:to:TzA}
\cM\lp\cI-\cM\rp \tz{Q}=\cI\tz{A},  
\end{equation}
where 
\begin{equation}
  \label{eq:def:TzA}
\tz{A}=\MR\lp\tz{I}-\MR\rp\lp\MR+\lb\tz{I}-\MR\rb\tz{S}\rp^{-1}=\Lt{a_1}{a_2}.
\end{equation}
Since $\cI$ is positively defined, the sign of \eqref{eq:LastTerm} is
determined by the sign of the dimensionless coefficients $a_1$ and
$a_2$. In Fig.~\ref{fig:TzA} we plot them for different values of
$\nI$, for the constraint from Eq.~\eqref{eq:ConstG}. These
coefficients are seen to be positive in the whole parameter range.
The corresponding graphs for constraints \eqref{eq:ConstGlobal} and
\eqref{eq:constraint:local} are similar and are not shown.

%


\end{document}